\documentclass[a4paper,12pt]{article}
\usepackage{enumerate}

\usepackage[svgnames]{xcolor}

\usepackage{geometry}
\usepackage{tikz}
\usepackage{amsmath,amsfonts,amsthm,amssymb,graphicx,natbib,geometry,arydshln,umoline,subfig,txfonts,setspace,enumerate}
\usepackage[title]{appendix}

\newtheorem{lemma}{Lemma}

\newtheorem{proposition}{Proposition}
\newtheorem{corollary}[proposition]{Corollary}
\newtheorem{claim}{Claim}

\theoremstyle{definition}

\theoremstyle{remark}

\begin{document}

\title{Strategic Ambiguity in Global 
Games\thanks{This paper builds on an earlier version circulated under the title ``Ambiguity and Risk in Global Games.'' 
I am most grateful to the associate editor and two anonymous reviewers for detailed comments and suggestions. 
I thank seminar participants for valuable discussions and comments at Bank of Japan, GRIPS, Hitotsubashi University, Hokkaido University, Kobe University, Nanzan University, the University of Tokyo, ZIF, PSE, UECE Lisbon Meetings 2009, CRETA - Marie Curie Conference 2010, Games 2016, and AMES 2016.  
I acknowledge financial support by MEXT, Grant-in-Aid for Scientific Research (grant numbers 20530150, 26245024, 18H05217).}}

\author{Takashi Ui\thanks{Email: \texttt{oui@econ.hit-u.ac.jp}. Postal address: HIAS, 2-1 Naka, Kunitachi, Tokyo 186-8601.}\\\small Hitotsubashi  University, Tokyo 186-8601, Japan\\
\small Kanagawa  University, Kanagawa 221-8686, Japan}

\date{June 2024}
\maketitle

\begin{abstract}
In games with incomplete and ambiguous information, rational behavior depends not only on fundamental ambiguity (ambiguity about states) but also on strategic ambiguity (ambiguity about others' actions), which further induces hierarchies of ambiguous beliefs.  
We study the impacts of strategic ambiguity in global games and demonstrate the distinct effects of ambiguous-quality and low-quality information. Ambiguous-quality information makes more players choose an action yielding a constant payoff, whereas (unambiguous) low-quality information makes more players choose an ex-ante best response to the uniform belief over the opponents' actions. If the ex-ante best-response action yields a constant payoff, sufficiently ambiguous-quality information induces a unique equilibrium, whereas sufficiently low-quality information generates multiple equilibria. In applications to financial crises, we show that news of more ambiguous quality triggers a debt rollover crisis, whereas news of less ambiguous quality triggers a currency crisis.\newline\noindent\textit{JEL classification numbers}: C72, D81, D82.
\end{abstract}

\newpage
\section{Introduction}

Consider an incomplete information game with players who have ambiguous beliefs about a payoff-relevant state.\footnote{See \citet{gilboamarinacci2013} and \citet{machinasiniscalchi2014} for a survey of the motivation and history of ambiguity and ambiguity aversion.} 
Players receive signals about a state, but they do not exactly know the true joint distribution of signals and a state.
In this game, players' beliefs about the opponents' actions are also ambiguous even if players know the opponents' strategies, which assign an action to each signal, because their beliefs about the opponents' signals are ambiguous. 
Thus, rational behavior depends not only on fundamental ambiguity (ambiguity about states) but also on strategic ambiguity (ambiguity about others' actions).\footnote{In the global game literature, uncertainty about others' actions is called strategic uncertainty \citep{morrisshin2002}, which is the source of belief hierarchies.} 
Strategic ambiguity, arising from fundamental ambiguity, further induces infinite hierarchies of ambiguous beliefs \citep{ahn2007}.

Hierarchies of ambiguous beliefs can have a substantial impact on equilibrium outcomes. To illustrate it, consider a continuum of players who decide whether or not to invest in a project. The payoff to ``investing'' is $2$ if the project succeeds and $-1$ if the project fails, while the payoff to ``not investing'' is $0$. 
The project's success depends upon the proportion of players to invest and a state, which is either $g$ (good) or $b$ (bad) with an equal probability $1/2$.  
The project succeeds if and only if the state is $g$ and more than two thirds of the players invest. 
If players have no additional information about the state, this game has two symmetric pure-strategy equilibria, where all players invest, and no players invest, respectively.

We now assume that each player receives a private signal about the state: he receives $G$ (resp.\ $B$) with probability $p\geq 1/2$ when the state is $g$ (resp.\ $b$), 
and signals are conditionally independent across players. 
We also assume that players do not know the true value of $p$ and make a decision on the basis of the most pessimistic assessment of $p$ conforming to maxmin expected utility (MEU) preferences  \citep{gilboaschmeidler1989} and prior-by-prior (full Bayesian) updating \citep{faginhalpern1990, jaffray1992, pires2002}. 
Then this game has a unique equilibrium where no players invest. 
To see why, consider first a player who receives signal $B$. 
The conditional probability of state $b$ is $p$ by Bayes rule, so the worst-case scenario for investing is that the state is bad with probability $1$. 
Facing such fundamental ambiguity, this player does not invest as if the signal were precise. 
Consider next a player who receives signal $G$. 
The expected proportion of the opponents with signal $B$ is $2p(1-p)\leq 1/2$ by Bayes rule and the law of large numbers, 
so the worst-case scenario for investing is that half of the opponents receive signal $B$ and do not invest. 
Facing such strategic ambiguity, this player does not invest as if the signal were imprecise.
To summarize, ambiguous signals with unknown $p\in [1/2,1]$ give rise to a unique rationalizable strategy,\footnote{In this example, players have a set of priors given by $p\in [1/2,1]$, but we can obtain the same conclusion by replacing $[1/2,1]$ with $[\underline{p},\overline{p}]$ such that $1/2\leq\underline{p}<1/2+\sqrt{3}/6$ and $2/3<\overline{p}\leq 1$.} wheres uninformative signals with known $p=1/2$ result in multiple equilibria. 

This paper studies the impact of strategic ambiguity in binary-action supermodular games of incomplete information in which players receive noisy private signals about a state, i.e., global games \citep{carlssonvandamme1993a}.\footnote{See the surveys in \citet{morrisshin2002} and \citet{angeletoslian2016}.} 
Our model is a global game equipped with multiple priors, where players make a decision conforming to MEU preferences and prior-by-prior updating. 
In contrast to the above example, only a tiny portion of players have a dominant action facing fundamental ambiguity. 
However, strategic ambiguity amplifies the effect of fundamental ambiguity through belief hierarchies on the opponents' dominant actions.
As a result, the role of ambiguous-quality  information can be quite different from that of low-quality information. 

We first show that our model is also a supermodular game in terms of MEU preferences and develop a tractable procedure to analyze it. Because players evaluate each action by the most pessimistic beliefs, they behave as if they adopted different priors to evaluate different actions. 
Thus, we can analyze the model using a fictitious game with a pair of priors, each of which is in the set of priors and separately assigned to each action. 
We show that if every fictitious game admits a unique equilibrium, then our model also admits a unique equilibrium that survives iterated deletion of strictly interim-dominated strategies in terms of MEU preferences. 

A fictitious game with a pair of priors reduces to a single-prior game if one of the actions yields a constant payoff, which is referred to as a safe action \citep{morrisshin2002}.  
Our model with a safe action has a unique equilibrium not only when every single-prior game has a unique equilibrium but also when some have multiple equilibria. 
The latter is the case if the following holds: information quality is sufficiently ambiguous, and a safe action is an ex-ante best response (a best response before receiving a signal) to the uniform belief over the opponents' actions, which is referred to as an ex-ante Laplacian action.
The unique equilibrium coincides with the equilibrium of the single-prior game that maximizes the range of signals to which the safe action is assigned, where the maximum is taken over the sets of all equilibria and all priors.

In summary, we find two effects of ambiguous information on equilibrium outcomes when one action is a safe action. 
First, ambiguous information enlarges the range of signals assigned to a safe action, whereas it is known that low-quality information enlarges the range of signals assigned to an ex-ante Laplacian action. 
Thus, if a safe action is not ex-ante Laplacian, the effect of ambiguous information is opposite to that of low-quality information.\footnote{\citet{kawagoeui2013} conducted a laboratory experiment using a two-player global game and obtained data supporting this result.} 
Even if a safe action is ex-ante Laplacian, the former is also distinct from the latter in another sense. 
In this case, sufficiently ambiguous information induces a unique equilibrium, whereas it is known that sufficiently low-quality information induces multiple equilibria. 
These effects of ambiguous information are attributed to ambiguous belief hierarchies and MEU preferences, which make a safe action more survivable in iterated deletion of strictly interim-dominated strategies. 
In contrast, a safe action does not play such a special role in standard global games because two actions' payoff differential determines best responses.

Our findings have the following implications for the question of whether ambiguous information contributes to financial crises originating from coordination failures, which can be modeled as global games of regime change. 
In the model of a currency crisis \citep{obstfeld1996,morrisshin1998}, speculators must decide whether to attack a currency, where a crisis occurs if sufficiently many speculators attack the currency. 
We find that news of more ambiguous quality decreases the likelihood of the crisis because not to attack is a safe action. 
In the model of a debt rollover crisis \citep{calvo1988,morrisshin2004}, creditors must decide whether to roll over a loan, where a crisis occurs if sufficiently many creditors do not roll over the loan. 
We find that news of more ambiguous quality increases the likelihood of the crisis because not to roll over is a safe action. 
These results complement and contrast with the findings of \citet{iachannenov2015} on the effects of low-quality information in standard global games of regime change. 
They show that news of lower quality increases the likelihood of a currency crisis, but it does not influence that of a debt rollover crisis,\footnote{This is true in the model of \citet{morrisshin2004}.  \citet{iachannenov2015} study another model of a debt rollover crisis, where news of lower quality influences the likelihood of the crisis.} which is in sharp contrast to the above effects of ambiguous information.

The rest of this paper is organized as follows. 
Section 2 illustrates our findings using an example. 
In Section 3, we set up our model and report the main results. 
In Section 4, we discuss applications to financial crises. 
Section 5 provides several extensions. 
The last section concludes the paper.

\subsection{Related literature}\label{Related literature}

This paper joins a growing literature on the theory and applications of incomplete information games with MEU players.  
There are many applications to auctions and mechanism design. 
Earlier papers examined equilibria of first price sealed bid auctions with MEU agents  \citep{saloweber1995,lo1998}.  
More recent papers study optimal mechanism design \citep{boseetal2006, bosedaripa2009, bodohcreed2012} and implementability of social choice functions 
\citep{wolitzky2016, song2018, decastroyannelis2018,guoyannelis2021}. 
Ambiguous beliefs are exogenously given in these papers, whereas \citet{boserenou2014} and \citet{ditilloetal2017} allow a mechanism designer to endogenously engineer ambiguity. 
Using ambiguous mechanisms,  \citet{boserenou2014} characterize implementable social choice functions, and \citet{ditilloetal2017} solve revenue maximization problems. 
Other recent applications include strategic voting \citep{ellis2016,pan2019,ryan2021,fabrizi2019}.


Because our focus is on global games with applications to financial crises, this paper is also related to a theoretical literature on financial crises associated with ambiguity. 
\citet{caballerokrishnamurthy2008} explain flight to quality episodes using a variant of the Diamond-Dybvig model \citep{diamonddybvig1983} with agents in the face of ambiguity about liquidity shocks. 
\citet{dicksfulghieri2019} propose a theory of systemic risk using another variant of the Diamond-Dybvig model with two banks investing in ambiguous assets. 
Our work contributes to the literature\footnote{Other papers include \citet{uhlig2010} and \citet{routledgezin2009}.} by presenting the global game approach to financial crises under ambiguity, which can also be applied to other global game models of financial crises.\footnote{See \citet[][Section 5]{angeletoslian2016} for a survey.} 


This paper builds on the following foundational studies. 
\citet{epsteinwang1996} construct hierarchies of general preferences, thus providing a foundation for rationalizability and iterated deletion of strictly interim-dominated strategies in terms of  MEU preferences \citep[cf.][]{epstein1997}. 
\citet{ahn2007} constructs a universal type space with multiple beliefs analogous to  \citet{mertenszamir1985}. 
\citet{kajiiui2005} introduce a general class of incomplete information games \citep{harsanyi1967} with MEU players 
and two interim equilibrium concepts;\footnote{Incomplete information games with more general preferences are studied by \citet{azrieliteper2011}, \citet{gratetal2016}, and \citet{hananyetal2020}.}   \citet{kajiiui2009} and \citet{martinsdarocha2010} study the corresponding ``agreement theorem'' \citep{aumann1976} and its converse. 
Our solution concept corresponds to  rationalizability in \citet{epsteinwang1996} and one of the solution concepts in 
\citet{kajiiui2005}.


\section{A linear example}\label{section linear example}

This section illustrates our main findings using a canonical linear-normal global game \citep{morrisshin2001}. 
A continuum of players have two actions, action 0 and action 1, 
which are interpreted as not investing and investing, respectively. 
The payoff to action 1 is $\theta + l -1$, where $\theta\in \mathbb{R}$ is normally distributed with mean $y \in (0,1)$ and precision $\eta>0$  (i.e.\ variance $1/\eta$), and $l\in [0,1]$ is the proportion of the opponents choosing action 1. 
The payoff to action 0 is a constant $0$, so we say that action 0 is a safe action. 
In summary, each player's payoff function is 
\[
u(a,l,\theta)
=
\begin{cases}
0&\text{ if } a=0\text{ (not investing)},\\\theta+l-1&\text{ if } a=1\text{ (investing)}.
\end{cases}
\]
Player $i$ observes a private signal $x_i=\theta+\varepsilon_i$, where a noise term $\varepsilon_i$ is independently normally distributed with mean $0$ and precision $\xi>0$ (i.e.\ variance $1/\xi$), which can be regarded as a measure of information quality.

If it is common knowledge that $0<\theta<1$ or if players receive no signals about $\theta$, this game has two symmetric pure-strategy equilibria. 
If the interim expected value of $\theta$ is strictly greater than $1$, action 1 is a dominant action; if that of $\theta$ is strictly less than $0$, action 0 is a dominant action. 
When a player has the uniform belief over the opponents' actions (i.e.\ the expected value of $l$ is $1/2$),  
the ex-ante best response (the best response before receiving a signal) is action 1 if $y\geq 1/2$ and action 0 if $y\leq 1/2$.
We call it an ex-ante Laplacian action.\footnote{The notion of an ex-ante Laplacian action is different from that of a Laplacian action in \citet{morrisshin2002}, which is the best response to the uniform belief over the opponents' actions when a player knows $\theta$. A Laplacian is action 1 if $\theta\geq 1/2$ and action 0 if $\theta\leq 1/2$.} 



We assume that information quality is ambiguous. 
Players have a set of priors indexed by a measure of information quality $\xi\in \Xi\equiv [\underline\xi,\overline\xi]\subset\mathbb{R}_{++}$, and make a decision based on the most pessimistic assessment conforming to maxmin expected utility (MEU) preferences  \citep{gilboaschmeidler1989} and prior-by-prior updating \citep{faginhalpern1990, jaffray1992, pires2002}.\footnote{Players are assumed to believe that the precision is the same for all players, but we can drop this assumption without qualitatively changing our results. See Section \ref{Heterogeneous information quality}.} 
Thus, after receiving a private signal, each player has a set of interim beliefs indexed by $\xi\in \Xi$ and evaluates each action in terms of the minimum expected payoff, where the minimum is taken over $\Xi$.\footnote{Players are assumed to have a common set of priors. See \citet{kajiiui2009} for an implication of this assumption.}

We denote this game by $(u,\Xi)$. 
Consider a monotone strategy, where a player chooses action 1 if and only if a private signal is above a cutoff point $\kappa\in \mathbb{R}\cup \{-\infty, \infty\}$. This strategy is referred to as a switching strategy (with cutoff $\kappa$) and denoted by $s[\kappa]:\mathbb{R}\to \{0,1\}$, i.e., 
\begin{equation}
s[\kappa](x)\equiv 
\begin{cases}
1 & \text{if } x>\kappa,\\
0 & \text{if } x\leq \kappa.
\end{cases}\label{def: s[k]}
\end{equation}
To characterize the best response to $s[\kappa]$, consider a player with a private signal $x$ who believes that the opponents follow $s[\kappa]$. 
The expected payoff to action 1 with respect to $\xi$ is calculated as 
\begin{equation}
\pi_{\xi}^1(x, \kappa)\equiv E_\xi[\theta|x]+\text{Prob}_\xi[x'>\kappa|x]-1
=\frac{\eta y+\xi x}{\eta+\xi}-
\Phi\left(\sqrt{\frac{\xi(\eta+\xi)}{\eta+2\xi}}\left(\kappa-\frac{\eta y+\xi x}{\eta+\xi}\right)\right), \label{liner expected payoff }
\end{equation}
where $\text{Prob}_\xi[x'>\kappa|x]$ is the probability that an opponent receives a private signal greater than $\kappa$, and $\Phi$ is the cumulative distribution function of the standard normal distribution. 
Then this player prefers action 1 to action 0 if and only if 
$
\min_{\xi \in \Xi} \pi^1_{\xi}(x, \kappa)\geq 0$. 
Thus, $s[\kappa']$ is a unique best response to $s[\kappa]$ 
if $\min_{\xi \in \Xi} \pi^1_{\xi}(\kappa', \kappa)= 0$ because $\min_{\xi \in \Xi} \pi^1_{\xi}(x, \kappa)$ is strictly increasing in $x$ and decreasing in $\kappa$. 
In particular, a strategy profile where all players follow $s[\kappa]$ is an equilibrium if $\kappa$ is a solution to 
\begin{equation}
\min_{\xi \in \Xi} \pi^1_{\xi}(\kappa, \kappa)= 0, \label{eq condition}
\end{equation}
which is referred to as a switching equilibrium (with cutoff $\kappa$). 
This implies that a unique switching equilibrium exists if (\ref{eq condition}) has a unique solution, and its strategy is shown to be a unique strategy surviving iterated deletion of strictly interim-dominated strategies with respect to MEU preferences by an argument similar to that in the standard global game analysis. 

\begin{figure} 
  \centering
  \subfloat[$y=0.52>1/2$]{\label{fig:y>1/2}\includegraphics[width=5.1cm, bb=0 0 360 223]{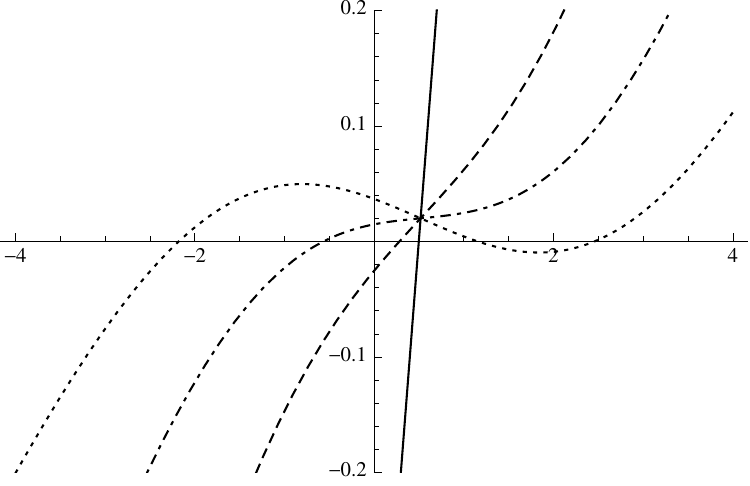}}\ 
  \subfloat[$y=0.48<1/2$]{\label{fig:y<1/2}
 \quad \includegraphics[width=5.1cm, bb=0 0 360 223]{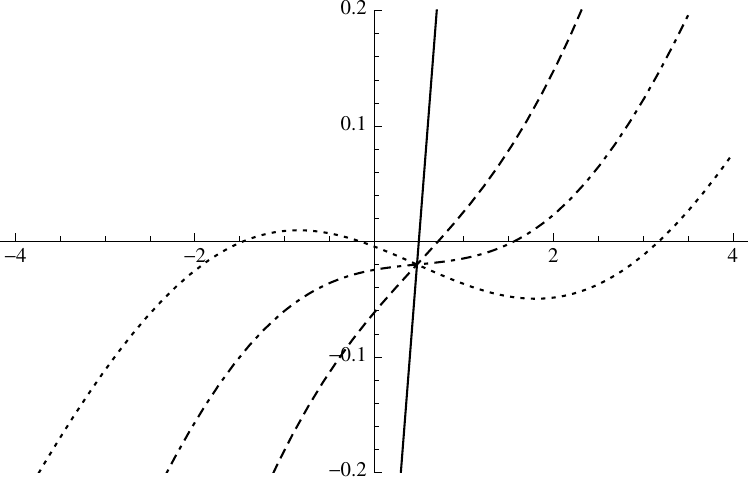}}
\caption{\small Graphs of $\pi^1_\xi(\kappa,\kappa)$ with $\eta=2$, $\xi=10^6$ in the solid line, $\xi=1.1$ in the dashed line, 
$\xi=0.56$ in the dash-dot line, and $\xi=0.37$ in the dotted line. 
}
  \label{fig1}
\end{figure}

An equilibrium of $(u,\Xi)$ is related  in several ways to an equilibrium of a single-prior game $(u,\{\xi\})$ with $\xi\in \Xi$. 
Most importantly, the maximum equilibrium cutoff in $(u,\Xi)$ 
\begin{align}
	k(\Xi)\equiv \max \{\kappa\mid \min_{\xi\in \Xi}\pi^1_{\xi}(\kappa, \kappa)= 0\}
\notag 
\end{align}
equals the maximum of the maximum equilibrium cutoffs in single-prior games over $\xi\in \Xi$, as stated in the following claim. This implies that ambiguous information increases the maximum equilibrium cutoff.  
\begin{claim}\label{claim 0}
It holds that 
\begin{align}
k(\Xi)=\max_{\xi\in\Xi}k(\xi)=\max_{\xi\in\Xi}\left(\max \{\kappa\mid \pi^1_{\xi}(\kappa, \kappa)= 0\}\right)
\label{ke}	
\end{align}
where we use $k(\xi)$ to denote $k(\{\xi\})$, the maximum equilibrium cutoff in $(u,\{\xi\})$, with some abuse of notation. 
\end{claim}
We can illustrate \eqref{ke} using Figure~\ref{fig1}, where a horizontal intercept in the graph of $\pi^1_\xi(\kappa,\kappa)$ is a solution to $\pi^1_{\xi}(\kappa, \kappa)= 0$. 
Consider $(u,\Xi)$ with $\Xi=[0.56,1.1]$. 
The graph of $\min_{\xi\in \Xi}\pi^1_\xi(\kappa,\kappa)$ is the lower envelope of the dashed line 
and the dash-dot line. 
Thus, the maximum equilibrium cutoff in $(u,\Xi)$ is the horizontal intercept of the dashed line in Figure \ref{fig:y>1/2} 
and that of the dash-dot line in Figure \ref{fig:y<1/2}. 
To see why \eqref{ke} holds more formally, note that if $\kappa>\max_{\xi\in\Xi}k(\xi)$, then $\pi_\xi^1(\kappa,\kappa)>0$ for all $\xi\in \Xi$ because $\lim_{\kappa\to\infty}\pi_\xi^1(\kappa,\kappa)=\infty$ by \eqref{liner expected payoff }. 
Thus, 
\[
\min_{\xi\in \Xi}\pi^1_\xi(\kappa,\kappa)
\begin{cases}
>0 &\text{ if }\kappa>\max_{\xi\in\Xi}k(\xi),\\
\leq \pi_{\hat\xi}^1(\kappa,\kappa)=0 &\text{ if }\kappa=\max_{\xi\in\Xi}k(\xi)
=k(\hat\xi)  
\end{cases}
\]
because $\Xi$ is compact and $\pi^1_\xi(\kappa,\kappa)$ is continuous in $\xi$, which means $\max_{\xi\in\Xi}k(\xi)=k(\Xi)$. 


In particular, 
if $(u,\Xi)$ has a unique switching equilibrium, its cutoff is $\max_{\xi\in \Xi}k(\xi)$. 
It is straightforward to show that 
$(u,\Xi)$ has a unique equilibrium if $(u,\{\xi\})$ has a unique equilibrium for each $\xi\in \Xi$, which is true if $\xi>\xi^*\equiv \eta\left(\eta-2\pi+({\eta^2+12\pi\eta+4\pi^2})^{1/2}\right)/(8\pi)$, as shown by \citet{morrisshin2001}. Thus, we obtain the following claim.

\begin{claim}\label{linear proposition 2}
If the minimum precision in $\Xi$ is strictly greater than $\xi^*$, 
there exists a unique switching equilibrium with cutoff $k(\Xi)=\max_{\xi\in\Xi}k(\xi)$.  
\end{claim}

\begin{figure} 
  \centering
  \subfloat[\mbox{Ambiguous quality $\Xi=[1,3]$}]{\label{fig:ambiguous}\includegraphics[width=5.1cm, bb=0 0 360 232]{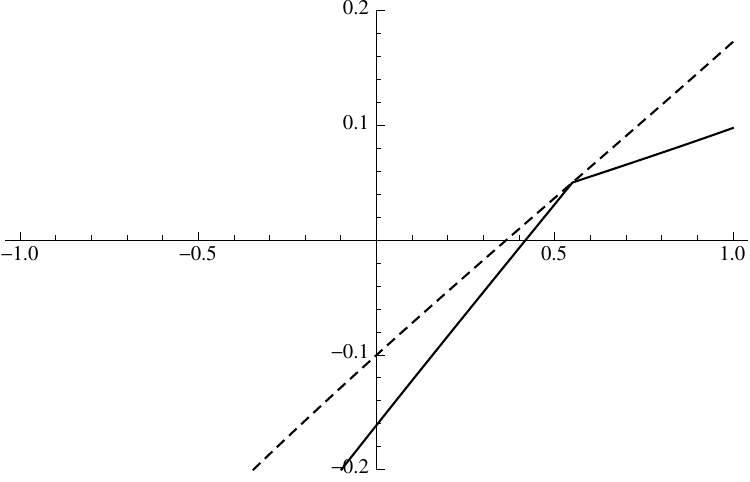}} 
  \subfloat[Low quality $\Xi=\{1\}$]{\label{fig:low}
  \quad \includegraphics[width=5.1cm, bb=0 0 360 232]{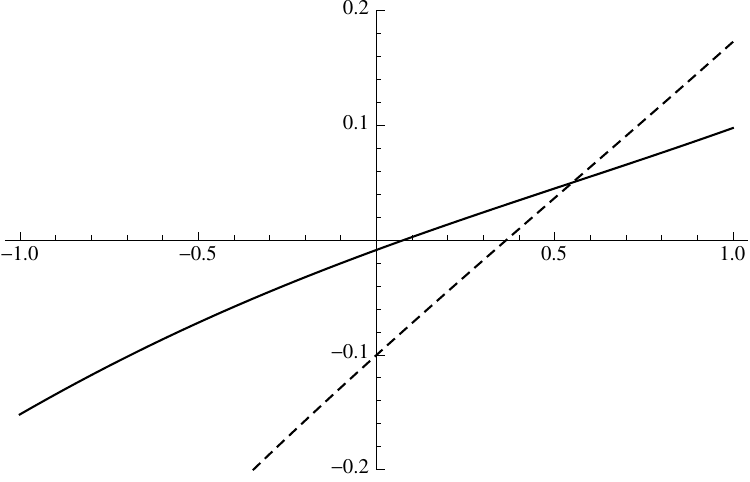}}\ 
\caption{\small Graphs of $\displaystyle\min_{\xi\in\Xi}\pi^1_\xi(\kappa,\kappa)$ with $y=0.55$, $\eta=2$, and $\Xi=\{2\}$ in the dashed line.}
  \label{fig2}
\end{figure}

If action 1 is ex-ante Laplacian (i.e., $y>1/2$),  ambiguous-quality information and low-quality information have different effects on the equilibrium cutoff by Claim \ref{linear proposition 2}. 
To illustrate this, consider the following three games: a benchmark game $(u,\{\xi^0\})$ with $\xi^0>\xi^*$, a game with low-quality information $(u,\{\xi^L\})$ with $\xi^0>\xi^L>\xi^*$, and a game with ambiguous-quality information $(u,[\underline{\xi},\overline{\xi}])$ with $\overline{\xi}>\xi^0>\underline{\xi}>\xi^*$. 
In a single-prior game, low-quality information enlarges the range of signals assigned to an ex-ante
Laplacian action (see Figure \ref{fig1}), as implied by the discussion of \citet{morrisshin2001}. 
Thus, low-quality information decreases the equilibrium cutoff, i.e., $k(\xi^L)<k(\xi^0)$ (see Figure \ref{fig:low}). Conversely, ambiguous-quality information increases the equilibrium cutoff, i.e., $k(\xi^0)< \max_{\xi\in [\underline{\xi},\overline{\xi}]}k(\xi)=k(\overline{\xi})$ \mbox{(see Figure~\ref{fig:ambiguous}).}


If action 0 is ex-ante Laplacian (i.e., $y<1/2$), the two types of information have similar effects on the equilibrium cutoff, but they have different effects on the number of equilibria. 
That is, $(u,\Xi)$ admits a unique equilibrium if the information quality is ambiguous enough, although $(u,\{\xi\})$ has multiple equilibria if the information quality is low enough. 

We demonstrate this informally using $(u,\{0\})$ and $(u,[0,\overline{\xi}])$ (in our formal analysis, the lowest precision is assumed to be strictly positive), which is analogous to the example in the introduction.
In $(u,\{0\})$, a player regards a private signal as uninformative, and both $s[-\infty]$ (investing for every signal) and $s[\infty]$ (not investing for any signal) are equilibrium strategies. 
In $(u,[0,\overline{\xi}])$, by contrast, a player regards a private signal as either informative or uninformative depending upon the worst-case scenario for investment. 
As in the case of $(u,\{0\})$, $s[\infty]$ is an equilibrium strategy because it is mutually optimal for each player not to invest under the scenario that a private signal is uninformative. 
However, $s[-\infty]$ is not an equilibrium strategy. This is because when all opponents invest and a player receives bad news about investment, 
the player prefers not to invest under the scenario that the bad news is reliable.\footnote{If a private signal $x$ is less than $-\eta y/\overline{\xi}$, the best response is not to invest   because $\min_{\xi\in \Xi}\pi^1_\xi(x,-\infty)< \pi^1_{\overline{\xi}}(-\eta y/\overline{\xi},-\infty)=0$.} 
Consequently, $s[\infty]$ is a unique equilibrium strategy in $(u,[0,\overline{\xi}])$.

Formally, we have the following claim.

\begin{claim}\label{new claim 3}
Suppose that $y<1/2$. 
There exists a unique switching equilibrium with cutoff $k(\Xi)=\max_{\xi\in\Xi}k(\xi)$ if either of the following conditions holds.
{\em (i)} For any minimum precision $\underline{\xi}>0$, the maximum precision $\overline{\xi}>0$ is sufficiently high.
{\em (ii)} For any maximum precision $\overline{\xi}>0$, the minimum precision $\underline{\xi}>0$ is sufficiently low. 
\end{claim}

\begin{figure} 
  \centering
{\label{fig:type I}\includegraphics[width=5.04cm, bb=0 0 360 232]{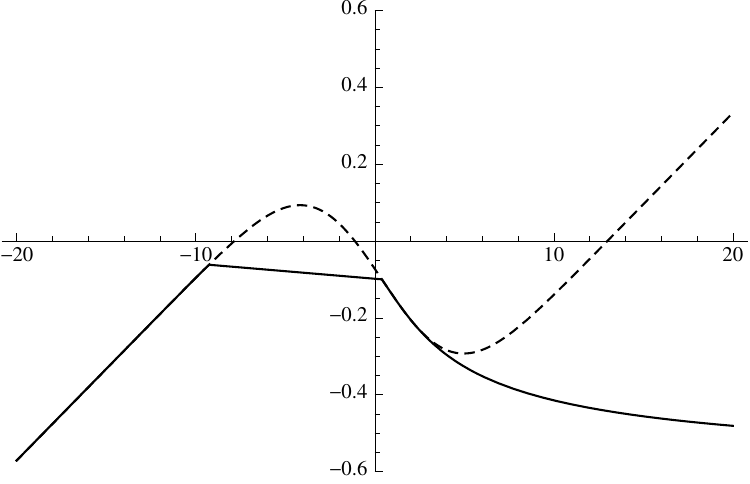}}
\caption{\small Graphs of $\displaystyle\min_{\xi\in\Xi}\pi^1_\xi(\kappa,\kappa)$ with $y=0.4$, $\eta=2$, $\Xi=[1/10^4,1/10]$ in the solid line, 
and $\Xi=\{1/10\}$ in the dashed line.}
  \label{fig3}
\end{figure}

We can illustrate (i) using Figure~\ref{fig:y<1/2}. Fix $\underline{\xi}=0.37$. 
When $\overline{\xi}=\underline{\xi}$, 
the graph of $\min_{\xi\in \Xi}\pi^1_\xi(\kappa,\kappa)$ is the dotted line,  
so $(u,\Xi)$ has three equilibria.  
In contrast, when $\overline{\xi}=0.56$, the graph of $\min_{\xi\in \Xi}\pi^1_\xi(\kappa,\kappa)$ is the lower envelope of the dash-dot line and the dotted line, so $(u,\Xi)$ has a unique equilibrium.  
We can also illustrate (ii) using Figure~\ref{fig3}. 
Fix $\overline{\xi}=1/10$. 
When $\underline{\xi}=\overline{\xi}$, 
the graph of $\min_{\xi\in \Xi}\pi^1_\xi(\kappa,\kappa)$ is the dashed line,  
so $(u,\Xi)$ has three equilibria.  
In contrast, when $\underline{\xi}=1/10^4$, the graph of $\min_{\xi\in \Xi}\pi^1_\xi(\kappa,\kappa)$ is the solid line, so $(u,\Xi)$ has a unique equilibrium with a very large cutoff.

Claim \ref{new claim 3} is based upon the following observation: (\ref{eq condition}) has no solution less than $-\eta y/\overline{\xi}$ and exactly one solution greater than $y$.\footnote{The former holds because $\min_{\xi\in\Xi}\pi^1_\xi(\kappa,\kappa)<(\eta y+\overline{\xi}\kappa)/(\eta+\overline{\xi})<0$ for $\kappa<-\eta y/\overline{\xi}$. The latter holds because $\pi^1_\xi(y,y)=y-1/2<0$ for all $\xi\in \Xi$, $\min_{\xi\in\Xi}\pi^1_\xi(\kappa,\kappa)>0$ for sufficiently large $\kappa$, and $d^2\pi^1_\xi(\kappa,\kappa)/d\kappa^2$ is strictly positive for all $\kappa> y$ and $\xi\in \Xi$.} 
This implies that an equilibrium is unique if (\ref{eq condition}) has no solution on the interval $[-\eta y/\overline{\xi},y]$, or equivalently, 
\begin{equation}
\text{$\max_{\kappa\in [-\eta y/\overline{\xi},y]}\min_{\xi\in \Xi}\pi^1_{\xi}(\kappa,\kappa)<0$.}
\label{condition new claim 2}
\end{equation}
The condition (i) implies (\ref{condition new claim 2}) because if $\overline{\xi}$ is sufficiently large (i.e., $\overline{\xi}>\xi^*$) then $\pi^1_{\overline{\xi}}(\kappa,\kappa)=0$ has a unique solution, which is greater than $y$, and thus $\min_{\xi\in \Xi}\pi^1_{\xi}(\kappa,\kappa)
\leq\pi^1_{\overline{\xi}}(\kappa,\kappa)<0$ for all $\kappa\in[-\eta y/\overline{\xi},y]$. 
The condition (ii) also implies (\ref{condition new claim 2}) because the left-hand side of \eqref{condition new claim 2} is less than $\pi^1_{\underline{\xi}}(y,-\eta y/\overline{\xi})$, which converges to $y-1/2<0$ as $\underline{\xi}$ approaches $0$.

As Figure~\ref{fig3} illustrates, $k(\Xi)$ can be arbitrarily large for sufficiently small $\underline{\xi}$. This is because $k(\Xi)$ equals the maximum equilibrium cutoff in the single-prior game with the lowest precision,\footnote{If $(u,\Xi)$ has a unique equilibrium, $k(\Xi)=k(\underline{\xi})$ if $y<1/2$ and $k(\Xi)=k(\overline{\xi})$ if $y>1/2$, which is illustrated in Figure \ref{fig1} and follows from the effect of low-quality information on an ex-ante Laplacian action.} and it goes to infinity as the lowest precision approaches zero: 
\[
k(\Xi)=\max_{\xi\in\Xi}k(\xi)=k(\underline{\xi})\to \infty \text{ as } \underline{\xi}\to 0.
\]
This suggests that $k(\Xi)$ is sensitive to $y$ when $\underline{\xi}$ is very small because $k(\Xi)$ is very large. 
To verify it, we apply the implicit function theorem to \eqref{eq condition} and obtain 
\begin{align*}
\frac{d k(\Xi)}{dy}=\frac{d k(\underline{\xi})}{dy}
=-\frac{{\eta }/({\eta +\underline{\xi} })+\gamma\phi(\gamma(y-k(\underline{\xi}) ))}{{\underline{\xi} }/({\eta +\underline{\xi} })-  \gamma\phi(\gamma(y-k(\underline{\xi}) ))}<-\frac{\eta }{\underline{\xi}}\to-\infty\text{ as }\underline{\xi}\to 0,
\end{align*}
where $\gamma=
\sqrt{{\eta^2\underline{\xi} }/({(\eta +\underline{\xi} ) (\eta +2 \underline{\xi} )})}$ and $\phi$ is the density function of the standard normal distribution. 
Thus, the unique equilibrium cutoff is decreasing in $y$, and the marginal decrease can be arbitrarily large for sufficiently small $\underline{\xi}$. 
Since $y$ is interpreted as a public signal, we can conclude that a small change in a public signal has a significant impact on the equilibrium cutoff when information quality is substantially ambiguous.


\section{Main results}\label{section main results}

In this section, we introduce a general model and present our main results. 
All proofs are relegated to the appendix.

There is a continuum of players indexed by $i\in [0,1]$.\footnote{We consider a continuum of players because many applications of global games assume so. It is straightforward to translate our results for a symmetric two-player case as in \citet{carlssonvandamme1993a}.} Each player has a binary action set $\{0,1\}$ and a payoff function $u:\{0,1\}\times [0,1] \times \mathbb{R}\to \mathbb{R}$, where $u(a,l,\theta)$ is a payoff to action $a\in \{0, 1\}$ when a proportion $l\in [0,1]$ of the opponents choose action 1 and a payoff state is $\theta\in \mathbb{R}$.  
An action $a\in \{0,1\}$ is called a safe action if $u(a,l,\theta)$ is independent of $(l,\theta)$. 

Player $i$ observes a noisy private signal $x_i=\theta+\varepsilon_i$, where $\theta$, $\varepsilon_i$, and all other $\varepsilon_j$'s are random variables that are mutually independent and follow a probability distribution unknown to the players. 
Each player with each private signal has MEU preferences determined by a payoff function $u$ and a set of beliefs indexed by a set of parameters $\Xi$. 
For each $\xi\in\Xi$, 
let $p_\xi(\theta)$, $q_\xi(\varepsilon_i)$, and  
$p_\xi(\theta|x_i)$ denote 
the probability density function of $\theta$, 
that of $\varepsilon_i$, 
and the conditional probability density function of $\theta$ given $x_i$, respectively.\footnote{Note that $p_\xi(\theta|x_i)=p_\xi(\theta) q_\xi(x_i-\theta)/\int p_\xi(\theta) q_\xi(x_i-\theta)dx_i$, but our discussion goes through even if $p_\xi(\theta|x_i)$ is not a conditional probability density function. }
Examples of $\xi$ include, but are not limited to, the variances and means of the relevant random variables.
We assume that $\Xi$ is a compact and connected space. 



A game is denoted by a pair $(u, \Xi)$. 
A strategy in $(u,\Xi)$ is a measurable mapping $\sigma: \mathbb{R}\to \{0,1\}$, which assigns an action to each private signal. 
To formally define an equilibrium concept, consider a player who believes that the opponents follow a strategy $\sigma$. 
Let $l_\xi(\theta)$ be the player's evaluation of the proportion of opponents taking action 1 when the state is $\theta$ based on a single prior indexed by $\xi$.
Then, the expected payoff to action $a\in \{0,1\}$ conditional on a private signal $x$ under the same prior is 
\[
E_\xi[u(a, l_\xi(\theta), \theta)|x]\equiv \int u(a, l_\xi(\theta), \theta)p_\xi(\theta|x)d\theta.    
\]
Thus, this player prefers action $a$ to action $a'$ if and only if 
\begin{align}
\min_{\xi\in\Xi}E_\xi[u(a, l_\xi(\theta), \theta)|x]
 \geq 
\min_{\xi\in\Xi} E_\xi[u(a', l_\xi(\theta), \theta)|x]. 
\label{preference}
\end{align}
We assume that 
\begin{equation}
l_\xi(\theta)=
 E_\xi[\sigma|\theta]\equiv \int \sigma(\theta+\varepsilon)q_\xi(\varepsilon)d\varepsilon=\int \sigma(x)q_\xi(x-\theta)dx,  
 \label{LLN1}	
\end{equation}
that is, the player evaluates the proportion of opponents taking action 1 as the conditional expected value of $\sigma$ given $\theta$, which follows the global game analysis by \citet{morrisshin2002}.\footnote{See Section \ref{SLLN sun} for a justification of this assumption.}
Under this assumption, 
a strategy profile in which all players follow $\sigma$ is an equilibrium if $\sigma(x)=a$ is a best response for almost all $x$ in the sense of \eqref{preference} with \eqref{LLN1} for $a'\neq a$. 
We focus on a pure-strategy equilibrium.\footnote{A pure-strategy equilibrium may not satisfy the requirement of a mixed-strategy equilibrium in incomplete information games with MEU players, as discussed in \citet{kajiiui2005}. 
See Section \ref{Mixed-strategy equilibria}.}

To characterize a switching equilibrium, where all players follow a switching strategy $s[\kappa]$ given by \eqref{def: s[k]}, 
let 
$$\pi_{\xi}^a(x, \kappa)\equiv 
E_\xi[u(a, E_\xi[s[\kappa]|\theta], \theta)|x]$$ denote the expected payoff to action $a\in\{0,1\}$ with respect to $\xi$ when a player receives a private signal $x$ and the opponents follow $s[\kappa]$. 
Then $s[\kappa]$ is an equilibrium strategy if and only if  
\begin{equation}
\min_{\xi \in \Xi}\pi_{\xi}^1(x, \kappa)-\min_{\xi \in \Xi}\pi_{\xi}^0(x, \kappa)
\begin{cases}
\geq 0 &\text{if } x>\kappa, \\
\leq 0 & \text{if } x\leq \kappa.
\end{cases}
\label{min min (2)}
\end{equation}
If the left-hand side of (\ref{min min (2)}) is continuous and increasing in $x$,  (\ref{min min (2)})  reduces to 
\begin{equation}
\min_{\xi \in \Xi}\pi_{\xi}^1(\kappa, \kappa)-\min_{\xi \in \Xi}\pi_{\xi}^0(\kappa, \kappa)=0. \label{min min 0}
\end{equation} 

We assume the following conditions on $(u,\Xi)$. 
\begin{description}
\item[A1 (Action Monotonicity)] For each $\theta$, $u(1, l, \theta)$  is increasing in $l$; $u(0, l, \theta)$ is decreasing in $l$. 
\item[A2 (State Monotonicity)] For each $l$, $u(1, l, \theta)$ is increasing in $\theta$; $u(0, l, \theta)$ is decreasing in $\theta$. 

\item[A3 (Stochastic Dominance)] 
For each fixed $\xi\in\Xi$,  
if $x>x'$, then $p_\xi(\theta|x)$ first-order stochastically dominates 
$p_\xi(\theta|x')$. 

\item[A4 (Continuity)] $\pi_{\xi}^a(x, \kappa)$ exists for all 
$(x,\kappa, \xi)\in \mathbb{R}\times \mathbb{R}\cup\{-\infty,\infty\}\times\Xi$, and 
the function $(x,\kappa, \xi)\mapsto \pi_{\xi}^a(x, \kappa)$ is continuous for each $a\in \{0,1\}$. 

\item[A5 (Limit Dominance)] There exist $\underline{\theta},\overline{\theta}\in \mathbb{R}$ satisfying 
\begin{gather*}
u(1, 1, \underline\theta)- u(0,1, \underline\theta)< 0,\   \lim_{x\to-\infty}\int^{\underline\theta}_{-\infty}
 p_\xi(\theta|x)d\theta=1\text{ for each $\xi$},\\
u(1, 0, \overline\theta)- u(0,0, \overline\theta)> 0,\  \lim_{x\to\infty}\int_{\overline\theta}^{\infty} p_\xi(\theta|x)d\theta=1\text{ for each $\xi$}.
\end{gather*}
\end{description}

By A1 and A2, the payoff to action 1 (resp.\ action 0) is increasing  (resp.\ decreasing) in the proportion of the opponents choosing action 1 and the state. 
In a single-prior game, it is enough to assume monotonicity of the payoff differential $u(1, l, \theta)-u(0, l, \theta)$ \citep[cf.][]{morrisshin2002}.
In a multiple-priors game, however, we need monotonicity of $u(1,l,\theta)$ and $u(0,l,\theta)$ separately because worst-case beliefs depend upon actions. 
A3 implies  that, for each fixed $\xi$, high signals convey good news about $\theta$. A sufficient condition for A3 is the monotone likelihood ratio property of the density function of a private signal $x_i$ conditional on $\theta$, i.e. $q_\xi(x_i-\theta)$, for each $\xi$ \citep{milgrom1981}.\footnote{For $\theta>\theta'$, $q_\xi(x_i-\theta)/q_\xi(x_i-\theta')$ is increasing in $x$.}  
For example, 
if $q_\xi$ is a density function of the normal, exponential, or uniform distribution,  among others, 
then A3 holds. 
A4 is a technical assumption, and it is satisfied if a payoff function is continuous and bounded. 
In some applications of global games, a payoff function is discontinuous or unbounded, yet A4 is satisfied in most cases. 
A5 requires that action 1 (resp.\ action 0) be a dominant strategy for sufficiently high (resp.\ low) signals.


Under these assumptions, $s[\kappa]$ is an equilibrium strategy if and only if $\kappa$ is a solution to {(\ref{min min 0})} because the following lemma implies the equivalence of (\ref{min min (2)}) and (\ref{min min 0}). 
\begin{lemma}\label{basic property of v}
The function $(x,\kappa)\mapsto \min_{\xi \in \Xi}\pi_{\xi}^1(x, \kappa)-\min_{\xi \in \Xi}\pi_{\xi}^0(x, \kappa)$ is 
continuous,  increasing in $x$, and decreasing in $\kappa$. 
\end{lemma}

Moreover,  $(u,\Xi)$ exhibits strategic complementarities in the following sense.

\begin{lemma}\label{SC}
If $\sigma(x)\geq \sigma'(x)$ for all $x\in \mathbb{R}$, then 
\begin{align*}
 \min_{\xi\in\Xi}E_\xi[u(1, E_\xi&[\sigma|\theta], \theta)|x]-\min_{\xi\in\Xi}E_\xi[u(0, E_\xi[\sigma|\theta], \theta)|x]\\
& \geq  \min_{\xi\in\Xi}E_\xi[u(1, E_\xi[\sigma'|\theta], \theta)|x]-\min_{\xi\in\Xi}E_\xi[u(0, E_\xi[\sigma'|\theta], \theta)|x]\ \text{ for all }x\in \mathbb{R}. 
\end{align*}
\end{lemma}

Therefore, the set of strategies surviving iterated deletion of strictly interim-dominated strategies includes the maximum and minimum elements, which are the maximum and minimum equilibrium strategies \citep{milgromroberts1990,vives1990,vanzandtvives2007}. 
In particular, this set is a singleton if and only if (\ref{min min 0}) has a unique solution.  

\begin{proposition}\label{lemma CvD} 
Let $\underline{\kappa},\overline{\kappa}\in \mathbb{R}$ be the minimum and maximum solutions to {\em(\ref{min min 0})}, respectively.  
Then  $s[\underline{\kappa}]$ and $s[\overline{\kappa}]$ are equilibrium strategies in $(u,\Xi)$, and $s[\overline{\kappa}](x)\leq \sigma(x)\leq s[\underline{\kappa}](x)$ for all  $x\in \mathbb{R}$ if a strategy $\sigma$ survives iterated deletion of strictly interim-dominated strategies.
In particular, 
$s[\kappa^*]$ is the (essentially)\footnote{An action of a player with a private signal $\kappa^*$ can be any action in an equilibrium strategy surviving iterated deletion, as in the standard global game analysis.} unique strategy surviving the iterated deletion 
if and only if $\kappa^*=\underline{\kappa}=\overline{\kappa}$.
\end{proposition}



To study what condition guarantees the uniqueness of equilibrium, we consider a fictitious game with a pair of priors indexed by $(\xi_0, \xi_1)\in\Xi\times\Xi$, where players evaluate actions 0 and 1 using $\xi_0$ and $\xi_1$, respectively.  
If each fictitious game has a unique switching equilibrium, $(u,\Xi)$ also has a unique switching equilibrium, and the equilibrium cutoff equals the ``maxmin'' of the equilibrium cutoffs in the fictitious games, as shown by the next proposition, which generalizes Claim \ref{linear proposition 2}.

\begin{proposition}\label{main proposition}
Suppose that there exists a unique value $\kappa=k(\xi_0, \xi_1)$ solving  
\begin{equation}
\pi_{\xi_1}^1(\kappa, \kappa)-\pi_{\xi_0}^0(\kappa, \kappa)=0\label{simple min min}
\end{equation} 
for each $(\xi_0,\xi_1)\in \Xi\times \Xi$, and that 
$k:\Xi\times\Xi\to\mathbb{R}$ is bounded.
Then $s[\kappa^*]$ 
is the (essentially) unique strategy surviving iterated deletion of strictly interim-dominated strategies, where 
\begin{equation}
\kappa^*= \min_{\xi_0\in \Xi}\max_{\xi_1\in \Xi}k(\xi_0, \xi_1)
=\max_{\xi_1\in \Xi}\min_{\xi_0\in \Xi}k(\xi_0, \xi_1). \label{simple min min 2}
\end{equation}
\end{proposition}

For example, suppose that $\pi_{\xi}^0(\kappa, \kappa)$ is independent of $\xi$, which is typically the case when action 0 is a safe action.\footnote{Even if no action is a safe action, $\pi_{\xi}^0(\kappa, \kappa)$ can be independent of $\xi$, which is the case in the bank run model of \citet{goldsteinpauzner2005}. See Section \ref{An action yielding a sate-independent payoff and a bank run}.}  
Then we can use a single-prior game $(u,\{\xi\})$ as a fictitious game because \eqref{simple min min} reduces to $\pi_{\xi_1}^1(\kappa, \kappa)-\pi_{\xi_1}^0(\kappa, \kappa)=0$. 
Hence, if each single-prior game with $\xi\in \Xi$ has a unique equilibrium, $(u,\Xi)$ also has a unique equilibrium, whose cutoff coincides with the maximum of the  equilibrium cutoffs of the single-prior games with $\xi\in \Xi$ by (\ref{simple min min 2}). 
As argued by \citet{carlssonvandamme1993a} and \citet{morrisshin2002}, each single-prior game has a unique equilibrium in the following two typical cases: the variance of $\varepsilon_i$ is sufficiently small, which is assumed in Claim \ref{linear proposition 2}, and the variance of $\theta$ is sufficiently large, which will be assumed in Section~\ref{applications}.


Assuming that $\pi_{\xi}^0(\kappa, \kappa)$ is independent of $\xi$, we 
conduct comparative static analysis on the equilibrium cutoff. The following proposition shows that the maximum equilibrium cutoff in $(u,\Xi)$ (possibly with multiple equilibria)
equals the maximum of the maximum equilibrium cutoffs in single-prior games over $\xi\in \Xi$. 
Thus, ambiguous information increases the maximum equilibrium cutoff, generalizing Claim~\ref{claim 0} (a similar result holds when $\pi_{\xi}^1(\kappa, \kappa)$ is independent of $\xi$).

\begin{proposition}\label{corollary safe}
Suppose that $c^0(\kappa)\equiv \pi_{\xi}^0(\kappa, \kappa)$ is independent of $\xi$. Then $\max \{\kappa\mid\pi_{\xi}^1(\kappa, \kappa)=c^0(\kappa)\}$ is the maximum equilibrium cutoff in $(u,\{\xi\})$, and its maximum over $\Xi$ denoted by $\kappa^0\equiv \max_{\xi\in\Xi} \, \max \{\kappa\mid\pi_{\xi}^1(\kappa, \kappa)=c^0(\kappa)\}$ is the maximum equilibrium cutoff in $(u,\Xi)$. 
\end{proposition}

An immediate corollary of this proposition gives another sufficient condition for the uniqueness of equilibrium. 

\begin{corollary}\label{corollary safe 2}
Let $\xi^0\in \Xi$ be such that $\pi_{\xi^0}^1(\kappa^0, \kappa^0)=c^0(\kappa^0)$ in Proposition \ref{corollary safe}. 
If $(u,\{\xi^0\})$ has a unique switching equilibrium,   
then $s[{\kappa}^0]$ is the (essentially) unique strategy surviving iterated deletion of strictly interim-dominated strategies in $(u,\Xi)$.
\end{corollary}

This corollary focuses on a single-prior game $(u,\{\xi^0\})$ that has the switching equilibrium whose cutoff is equal to the maximum equilibrium cutoff in $(u,\Xi)$. It shows that if this is a unique equilibrium in $(u,\{\xi^0\})$, then it is also a unique equilibrium in $(u,\Xi)$. 

Finally, we demonstrate that sufficiently ambiguous information can induce a unique equilibrium. 
Let $\Xi\equiv [\underline\xi,\overline\xi]\subset \mathbb{R}_{++}$, where $\xi\in \Xi$ is the precision of a noise term in a private signal. 
Assume that the probability density function of a noise term $q_\xi(\varepsilon_i)$ has a support $\mathbb{R}$ and satisfies 
$\int \varepsilon_i q_\xi(\varepsilon_i)d\varepsilon_i=0$, $\int \varepsilon_i^2 q_\xi(\varepsilon_i)d\varepsilon_i=1/\xi$, $q_\xi(\varepsilon_i)=q_\xi(-\varepsilon_i)$, and $q_\xi(\varepsilon_i)\leq q_\xi(\varepsilon_i')$ if $\varepsilon_i\leq\varepsilon_i'<0$. 
Let action 0 be a safe and ex-ante Laplacian action; that is, 
$u(0,l,\theta)=c^0\in\mathbb{R}$ for each $l$ and $\theta$ and
$c^0> \int_{-\infty}^\infty u(1,1/2,\theta) p(\theta)d\theta$, where $\int_{-\infty}^\infty u(1,1/2,\theta) p(\theta)d\theta$ is the ex-ante expected payoff to action 1 when a player has the uniform belief over the opponents' actions. 
The next proposition shows that the minimum equilibrium cutoff can be arbitrarily large if the minimum precision $\underline{\xi}$ is sufficiently small, and there exists a unique equilibrium if a single-prior game with sufficiently small $\xi$ has at most one switching equilibrium with a sufficiently large cutoff.  

\begin{proposition}\label{ex-ante Laplacian proposition}
Consider $(u, [\underline\xi,\overline\xi])$ described above with arbitrarily fixed $\overline{\xi}>0$. 
Suppose that action 0 is a safe and ex-ante Laplacian action. 
Then the following holds. 
\begin{itemize}
\item For every ${\overline{\kappa}}\in \mathbb{R}$, there exists $\delta(\overline{\kappa})>0$ such that, if $\underline{\xi}\leq\delta(\overline{\kappa})$, then the minimum equilibrium cutoff in $(u, [\underline\xi,\overline\xi])$ is greater than ${\overline{\kappa}}$. 
\item Suppose that there exists $\overline{\kappa}\in \mathbb{R}$ such that a single-prior game $(u, \{\xi\})$ has at most one switching equilibrium with a cutoff greater than ${\overline{\kappa}}$ for every ${\xi}\leq\delta(\overline{\kappa})$. 
Then there exists $\delta'\in (0,\delta(\overline{\kappa}))$ such that, if $\underline{\xi}\leq\delta'$, then $s[{\kappa}^0]$ is the (essentially) unique strategy surviving iterated deletion of strictly interim-dominated strategies in $(u,[\underline\xi,\overline\xi])$, where $\kappa^0$ is given in Proposition \ref{corollary safe}. 
\end{itemize}
\end{proposition}

A special case is the second half of Claim~\ref{new claim 3}: $(u,[\underline{\xi},\overline{\xi}])$ has a unique equilibrium for any $\overline{\xi}$ if 
$\underline{\xi}$ is small enough. 
This is because action 0 is a safe and ex-ante Laplacian action, and for any $\xi>0$, $(u,\{\xi\})$ has exactly one switching equilibrium with a cutoff greater than $y$. 


\section{Applications to regime change}\label{applications}

We apply our main results to global games of regime change \citep{angeletosetal2007}, which have applications to financial crises such as currency crises \citep{morrisshin1998}, debt rollover crises \citep{morrisshin2004}, and bank runs \citep{goldsteinpauzner2005},  among others. 
We study the effect of ambiguous-quality information on the likelihood of currency crises and debt rollover crises and compare it with that of low-quality information studied by \citet{iachannenov2015}.

In a game of regime change, either regime 0 or 1 is realized depending upon a state $\theta$ and the proportion of players choosing action 0. 
More specifically, regime 0 occurs if and only if the proportion of players choosing action 0 is greater than $\theta$. The payoff differential $u(1,l,\theta)-u(0,l,\theta)$ equals $d_R(\theta)$ if regime $R\in \{0,1\}$ occurs: 
\[
u(1,l,\theta)-u(0,l,\theta)=
\begin{cases}
d_0(\theta) & \text{if }1-l\geq  \theta,\\
d_1(\theta) & \text{if }1-l < \theta,
\end{cases}
\]
where $d_0(\theta)$ and $d_1(\theta)$ are non-decreasing in $\theta$, bounded, and satisfies  $d_0(\theta)<0< d_1(\theta)$. 
Later, we will assume that either action 0 or 1 is a safe action. 

A state $\theta$ is drawn from the improper uniform distribution over the real line, and a noise term $\varepsilon_i$ is drawn from a normal distribution with mean zero and precision $\xi$. 
Then $p_\xi(\theta|x)=\sqrt{\xi}\phi(\sqrt{\xi}(\theta-x))$ and $q_\xi(\varepsilon)=\sqrt{\xi}\phi(\sqrt{\xi}\varepsilon)$, where $\phi$ is the probability density function of the standard normal distribution. 
When the state is $\theta$ and each player follows $s[\kappa]$, 
the proportion of players choosing action 0 is $\Phi(\sqrt{\xi}(\kappa-\theta))$, where $\Phi$ is the cumulative distribution function of the standard normal distribution. 
Let $\theta=\theta^*(\kappa,\xi)\in (0,1)$ be the unique 
solution to 
\begin{equation}
\Phi(\sqrt{\xi}(\kappa-\theta))= \theta.
\label{RC1}\end{equation}
Then regime 0 occurs if and only if $\theta\leq\theta^*(\kappa,\xi)$, so $\theta^*(\kappa,\xi)$ is referred to as a regime-change cutoff. 
We can readily show that $\theta^*(\kappa,\xi)$ is strictly increasing in $\kappa$ because regime 0 is more likely to occur when more players choose action 0.

A single-prior game $(u,\{\xi\})$ is the global game of regime change in \citet{iachannenov2015}, and it admits a unique switching equilibrium with cutoff $\kappa=k(\xi)$, which is the unique solution to 
\begin{equation}
\pi_{\xi}^1(\kappa, \kappa)-\pi_{\xi}^0(\kappa, \kappa)
=\int_{-\infty}^{\theta^*(\kappa,\xi)} d_0(\theta)\sqrt{\xi}\phi(\sqrt{\xi}(\theta-\kappa))d\theta+\int^{\infty}_{\theta^*(\kappa,\xi)} d_1(\theta)\sqrt{\xi}\phi(\sqrt{\xi}(\theta-\kappa))d\theta=0.\notag\label{RC2}
\end{equation}
As a special case, assume that $d_0(\theta)=d_0$ and $d_1(\theta)=d_1$ are constant. 
Then this equation reduces to 
\begin{equation}
d_0\Phi(\sqrt{\xi}(\theta^*(\kappa,\xi)-\kappa))+d_1(1-\Phi(\sqrt{\xi}(\theta^*(\kappa,\xi)-\kappa)))
=d_0+(d_1-d_0)\theta^*(\kappa,\xi)=0 \label{cutoff under constant differential}
\end{equation}
by (\ref{RC1}). 
Therefore, $\theta^*(\kappa,\xi)=d_0/(d_0-d_1)$ and $k(\xi)=d_0/(d_0-d_1)+\Phi^{-1}(d_0/(d_0-d_1))/\sqrt{\xi}$. 

We consider $(u,\Xi)$ with $\Xi=[\underline{\xi},\overline{\xi}]\subset \mathbb{R}_{++}$ assuming that one of the actions is a safe action.\footnote{We can relax this assumption. See Section \ref{An action yielding a sate-independent payoff and a bank run}.}
It is readily shown that $(u,\Xi)$ satisfies the condition in Proposition~\ref{main proposition}. Thus, $(u,\Xi)$ has a unique switching equilibrium with the following cutoff. 
\begin{equation}
\kappa^*=
\begin{cases}
\max_{\xi \in \Xi}k(\xi) &\text{ if action 0 is a safe action},\\ \min_{\xi \in \Xi}k(\xi) & \text{ if action 1 is a safe action}.
\end{cases}\label{regime change cutoff}	
\end{equation}
Therefore, we obtain the following comparative static results.
\begin{proposition}\label{regime cutoff action 0 1}
If action 0 is a safe action, the regime-change cutoff in $(u,\Xi)$ is
\begin{equation}
\theta^*(\kappa^*,\xi)= \theta^*(\max_{\xi' \in \Xi}k(\xi'),\xi)
=\max_{\xi' \in \Xi}\theta^*(k(\xi'),\xi)\geq \theta^*(k(\xi),\xi)\text{ for each $\xi\in \Xi$}. 
\notag
\end{equation}
Thus, ambiguous-quality information increases the regime-change cutoff and makes 
regime 0 more likely.  
If action 1 is a safe action, the regime-change cutoff in $(u,\Xi)$ is
\begin{equation}
\theta^*(\kappa^*,\xi)= \theta^*(\min_{\xi' \in \Xi}k(\xi'),\xi)
=\min_{\xi' \in \Xi}\theta^*(k(\xi'),\xi)\leq \theta^*(k(\xi),\xi)\text{ for each $\xi\in \Xi$}.
\notag
\end{equation}
Thus, ambiguous-quality information decreases the regime-change cutoff and makes 
regime 1 more likely.
\end{proposition}



To compare the effects of ambiguous-quality information and low-quality information, we use the result of \citet{iachannenov2015} on the effect of low-quality information, which is determined by the sensitivity of $d_0(\theta)$ and $d_1(\theta)$ with respect to $\theta$.\footnote{\citet{iachannenov2015} obtain more general results without assuming that either $d_0(\theta)$ or $d_1(\theta)$ is constant.}
\begin{proposition}
\label{low quality lemma}
If $d_0(\theta)$ is constant, $\theta^*(k(\xi),\xi)$ is increasing in $\xi$. 
If $d_1(\theta)$ is constant, $\theta^*(k(\xi),\xi)$ is decreasing in $\xi$. 
If $d_0(\theta)$ and $d_1(\theta)$ are constant, $\theta^*(k(\xi),\xi)$ is independent of $\xi$. 
\end{proposition}

From the above two propositions, we obtain the following corollary, which explains when ambiguous-quality information and low-quality information have opposite effects on the regime-change cutoff.

\begin{corollary}\label{regime-change corollary}
If $d_0(\theta)$ \emph{(\emph{resp.}\ $d_1(\theta)$)} is constant and action 0 \emph{(\emph{resp.\ action 1})} is a safe action, low-quality information decreases \emph{(\emph{resp.\ increases})} the regime-change cutoff, whereas ambiguous-quality information increases \emph{(\emph{resp.\ decreases})} it. 
\end{corollary}




\subsection*{Currency crises}\label{section currency crises}

We study the effect of ambiguous-quality information on the likelihood of currency crises using the currency attack model of \citet{morrisshin1998}. 
Speculators decide whether to attack the currency by selling it short. 
The current value of the currency is $e^*$. If an attack is successful, the currency collapses and  floats to the shadow rate $f(\theta)<e^*$, where $\theta$ is the state of fundamentals and $f(\theta)$ is increasing in $\theta$. 
There is a fixed transaction cost $t\in (0,e^*-f(\theta))$ of attacking. 
Thus, the net payoff to a successful attack is $e^*-f(\theta)-t$, while that to an unsuccessful attack is $-t$. 
An attack is successful if and only if the proportion of speculators attacking the currency is greater than $\theta$. 
Writing 1 for the action not to attack (and 0 for that to attack), we have the following payoff function: 
\begin{align}
u(a,l,\theta)&=
\begin{cases}
0&\text{if }a=1, \\
e^*-f(\theta)-t&\text{if $a=0$ and $1-l\geq \theta$},\\
-t&\text{if $a=0$ and $1-l< \theta$},
\end{cases}\notag
\end{align}
where $l$ is the proportion of the opponents not attacking the currency.  

Consider a game of regime change with the above payoff function, 
 where $d_0(\theta)=-(e^*-f(\theta)-t)$, $d_1(\theta)=t$, and  a currency crisis corresponds to regime 0. 
By Corollary \ref{regime-change corollary}, 
ambiguous-quality information and low-quality information have opposite effects on the likelihood of a currency crisis.
That is, ambiguous-quality information decreases the regime-change cutoff and makes a currency crisis less likely. 
In contrast,  low-quality information increases the  regime-change cutoff and makes a currency crisis more likely.

Note that if $\theta$ is within a certain range, 
a currency crisis does not occur under sufficiently ambiguous information, although it does occur under unambiguous information. 
This is because speculators may not attack the currency even when they receive bad news about the state of fundamentals,  considering the worst-case scenario that the information is unreliable and the state of fundamentals is not so bad. 

With this in mind, let us briefly review the events leading up to the 1997 Thai baht crisis \citep{lipsky1998,siamwalla2005}. The risk of devaluation was widely recognized at least a year before the crisis, and the first wave of attacks was made in November 1996, followed by two more. During this period, the Thai central bank used foreign exchange reserves to defend the baht, but hid the loss of reserves by selling dollars in the forward market. After the third wave of attacks in May 1997, the central bank ordered commercial banks to stop lending to non-residents, creating a squeeze on foreign speculators selling the baht short. Then there was a run in June by Thais, not foreign speculators, and the baht floated on July 2. 

The level of foreign exchange reserves is an important factor in the state of fundamentals, but speculators did not know the hidden true level. Thus, speculators' information about the state was ambiguous, which results in fewer speculators attacking the currency in our model.\footnote{In this section, information quality is assumed to be ambiguous, but Proposition \ref{regime cutoff action 0 1} is valid for any type of ambiguity, which follows from  Proposition~\ref{main proposition}.} This observation gives the counterfactual theoretical prediction that if the loss of reserves had not been hidden, more speculators would have attacked the currency, and the baht would have floated much earlier. This is also an intuitive prediction, which the Thai central bank would have been aware of.



\subsection*{Debt rollover crises}

We study the effect of ambiguous-quality information on the likelihood of debt rollover crises using the creditor coordination model of \citet{morrisshin2004}. 
Creditors hold a loan secured on collateral and decide whether to roll over the loan. 
A creditor who rolls over the loan receives $1$ if an underlying investment project succeeds and receives $0$ if the project fails. 
A creditor who does not roll over the loan receives the value of the collateral $\lambda\in(0,1)$. 
The project succeeds if and only if the proportion of creditors not rolling over the loan is less than $\theta\in \mathbb{R}$. 
Writing 1 for the action to roll over (and 0 for that not to roll over), we have the following payoff function: 
\begin{align}
u(a,l,\theta)&=
\begin{cases}
\lambda &\text{if } a=0,\\
1 &\text{if } a=1 \text{ and }1-l< \theta,\\
0 &\text{if } a=1 \text{ and }1-l\geq \theta,
\end{cases}\notag
\end{align}
where $l$ is the proportion of creditors rolling over the loan. 

Consider a game of regime change with the above payoff function, where $d_0(\theta)=-\lambda$,  $d_1(\theta)=1-\lambda$, and a debt rollover crisis corresponds to regime 0. 
Ambiguous-quality information increases the regime-change cutoff and makes a debt rollover crisis more likely because action 0 is a safe action.  
In a single-prior game, however, information quality has no effect on the regime-change cutoff, which is a constant $\lambda$ by \eqref{cutoff under constant differential}.

We examine which type of ambiguity leads to a debt rollover crisis. 
The regime-change cutoff in $(u,\Xi)$ is calculated from Proposition \ref{regime cutoff action 0 1}: 
\begin{equation}
\max_{\xi' \in \Xi}\theta^*(k(\xi'),\xi)=
\max_{\xi' \in \Xi}\theta^*(\lambda+\Phi^{-1}(\lambda)/{\xi'}^{1/2}
,\xi)=
\begin{cases}
\theta^*(\lambda+\Phi^{-1}(\lambda)/\overline{\xi}^{1/2},\xi)&\text{if }\lambda < 1/2,\\ 
\theta^*(\lambda+\Phi^{-1}(\lambda)/\underline{\xi}^{1/2},\xi)&\text{if }\lambda>1/2,
\end{cases}\notag\label{rchcuttoff}
\end{equation}
where we use the fact that $\Phi^{-1}(\lambda)\gtrless 0$ if $\lambda\gtrless 1/2$.
Thus, if $\lambda<1/2$, the regime-change cutoff is increasing in the maximum precision $\overline{\xi}$, and if $\lambda>1/2$, it is decreasing in the minimum precision $\underline{\xi}$, which leads us to the following proposition. 
\begin{proposition}\label{crisis proposition}
Suppose that $\lambda<\theta<\overline{\theta}(\lambda,\xi)$ and $\lambda\neq 1/2$, where 
\begin{align}
	\overline{\theta}(\lambda,\xi)\equiv 
\sup_{\Xi\subset \mathbb{R}_{++}}\max_{\xi' \in \Xi}\theta^*(k(\xi'),\xi)=
\begin{cases}
\theta^*(\lambda,\xi) &\text{ if }\lambda<1/2\\
1 &\text{ if }\lambda>1/2
\end{cases}\notag\label{supcut}
\end{align}
is the supremum of the regime-change cutoffs and $\xi$ is the true precision.  
Then a debt rollover crisis occurs in $(u,\Xi)$ if and only if either {\em (i)} $\lambda<1/2$ and $\overline{\xi}$ is sufficiently large, or {\em (ii)} $\lambda>1/2$ and $\underline{\xi}$ is sufficiently small.
\end{proposition}

The type of ambiguity that results in the crisis depends upon the value of the collateral~$\lambda$. 
To see why, consider the case of $\lambda<1/2$, for example. An ex-ante Laplacian action is action~1 (to roll over) in this case, which means that low-quality information enlarges the range of signals assigned to action 1. 
Thus, $k(\xi)$ is increasing in $\xi$, and $\max_{\xi\in \Xi}k(\xi)=k(\overline{\xi})$ is increasing in $\overline{\xi}$. 
Therefore, the crisis occurs if $\overline{\xi}$ is large enough.
Similarly, in the case of $\lambda>1/2$, action 0 (not to roll over) is ex-ante Laplacian, so the crisis occurs if $\underline{\xi}$ is small enough. 
Consequently, to prevent a debt rollover crisis when $\lambda<\theta<\overline{\theta}(\lambda,\xi)$, it is important to convince players that (i) information cannot be so precise if $\lambda<1/2$, (ii) information cannot be so imprecise if $\lambda>1/2$, or (iii) information quality is unambiguous.

\section{Discussion}\label{section discussion}


\subsection{An action yielding a sate-independent payoff and a bank run}\label{An action yielding a sate-independent payoff and a bank run}

In Section \ref{applications}, we assume that one of the actions is a safe action and $\theta$ is drawn from the improper uniform distribution.  
We can obtain the same results under a slightly weaker requirement. 
Suppose that action $a\in \{0,1\}$ yields a state-independent payoff; that is, $u(a,l,\theta)=f(l)$, where $f:[0,1]\to \mathbb{R}$. 
Then  $\pi_\xi^a(\kappa,\kappa)$ is independent of $\xi$ and $\kappa$ because 
\begin{align*}
\pi_\xi^a(\kappa, \kappa)&= \int_{-\infty}^\infty u(a, 1-Q_\xi(\kappa-\theta), \theta)q_\xi(\kappa-\theta)d\theta 
= \int_0^1 u(a, l, \kappa-Q_\xi^{-1}(1-l))dl
=\int_0^1 f(l)dl, 
\end{align*} 
where $Q_\xi(\kappa-\theta)=\Phi(\sqrt{\xi}(\kappa-\theta))$ and $q_\xi(\kappa-\theta)=\sqrt{\xi}\phi(\sqrt{\xi}(\kappa-\theta))$. 
Thus, the discussion in Section \ref{applications} remains valid even if a safe action is replaced with an action with a state-independent payoff.

An example with such an action is the global game model of bank runs studied by \citet{goldsteinpauzner2005}, which is a variant of the Diamond-Dybvig model with noisy private signals.  
Depositors must decide whether to withdraw money from a bank in period 1 or wait until period 2. 
If a depositor withdraws money in period 1, the bank pays a fixed amount of money to him until it runs out of money, following a sequential-service constraint. 
Thus, the payoff to early withdrawal is determined solely by depositors' decisions and independent of a state. 
An argument similar to that in Section \ref{applications} shows that ambiguous-quality information makes a bank run more likely. 
This result also complements that of \citet{iachannenov2015}, who show that low-quality information makes a bank run less likely. 
These results together imply that a bank run is less likely under unambiguous low-quality information.

\subsection{Mixed-strategy equilibria}\label{Mixed-strategy equilibria}

Players are assumed to use pure strategies, which is common in the ambiguity literature. 
However, we can also consider mixed strategies.  
Let $\bar\sigma:\mathbb{R}\to [0,1]$ denote a mixed strategy that assigns action 1 to a private signal $x$ with probability $\bar\sigma(x)$.  
\citet{kajiiui2005} consider two equilibrium concepts, a mixed-strategy equilibrium and an equilibrium in belief.  
A strategy profile in which all players follow $\bar\sigma$ is a mixed-strategy equilibrium if the following condition is satisfied: when a player receives a private signal $x$ and the opponents follow $\bar\sigma$, the minimum expected payoff to the mixed action $\bar\sigma(x)$ is greater than or equal to that to any other mixed action. 
The same strategy profile is an equilibrium in belief if the following condition is satisfied:  when a player receives a private signal $x$ and the opponents follow $\bar\sigma$, if $\bar\sigma(x)>0$, then the minimum expected payoff to action 1 is greater than or equal to that to action 0, and if $\bar\sigma(x)<1$,   then the minimum expected payoff to action 0 is greater than or equal to that to action 1. 

A pure-strategy equilibrium in this paper satisfies the condition of  an equilibrium in belief, but it may not satisfy that of a mixed-strategy equilibrium because ambiguity-averse players may prefer objective randomization.\footnote{See Section 4.3 of \citet{kajiiui2005}.  This property plays an essential role in \citet{ellis2016}.}
However, if one of the actions is a safe action, a pure-strategy equilibrium is also a mixed strategy-equilibrium. 
To see this, suppose that action 0 is a safe action with $u(0,l,\theta)=c_0\in \mathbb{R}$. 
If $\sigma:\mathbb{R}\to\{0,1\}$ is a pure-strategy equilibrium, then  
$\min_{\xi\in\Xi}E_\xi[u(\sigma(x), E_\xi[\sigma|\theta], \theta)|x]
 \geq  \min_{\xi\in\Xi}pE_\xi[u(1, E_\xi[\sigma|\theta], \theta)|x]+(1-p)c_0$  
for all $x\in \mathbb{R}$ and $p\in [0,1]$, so $\sigma$ is also a mixed-strategy equilibrium. 
This implies that the discussions in Sections \ref{section linear example} and \ref{applications} and the propositions except Proposition \ref{lemma CvD} remain valid even if we adopt a mixed-strategy equilibrium as the equilibrium concept.  

\subsection{Heterogeneous information quality}\label{Heterogeneous information quality}

In Sections \ref{section linear example} and \ref{applications}, each player is assumed to believe that the probability distribution of a private signal is the same for all the players. 
\citet{laskar2014} makes this observation based upon an earlier version of this paper \citep{ui2009}\footnote{This contains the linear example and the models of financial crises together with some of the main results.} and studies a linear-normal global game with two MEU players whose private signals may follow different probability distributions, where the expected value of a private signal is ambiguous.

The general model in Section \ref{section main results} allows the probability distributions to be different across players, for which the main results remain valid. 
For example, we can consider a game of regime change $(u,\Xi)$ with $\Xi=\Xi_1\times\Xi_2$, 
where $\xi_1\in \Xi_1$ and $\xi_2\in \Xi_2$ are the precision of a noise term in a player's private signal and that in his opponents' private signals, respectively. 
When the state is $\theta$ and each opponent follows $s[\kappa]$ in $(u,\{\xi\})$ with $\xi=(\xi_1,\xi_2)$, the proportion of the opponents choosing action 0 is $\Phi(\sqrt{\xi_2}(\kappa-\theta))$, so the regime-change cutoff in $(u,\{\xi\})$ is $\theta^*(\kappa,\xi_2)$ by \eqref{RC1}. 
Thus, the equilibrium cutoff in $(u,\{\xi\})$ solves 
\begin{equation}
\pi_{\xi}^1(\kappa, \kappa)-\pi_{\xi}^0(\kappa, \kappa)
=\int_{-\infty}^{\theta^*(\kappa,\xi_2)} d_0(\theta)\sqrt{\xi_1}\phi(\sqrt{\xi_1}(\theta-\kappa))d\theta+\int^{\infty}_{\theta^*(\kappa,\xi_2)} d_1(\theta)\sqrt{\xi_1}\phi(\sqrt{\xi_1}(\theta-\kappa))d\theta=0 \notag
\end{equation}
because $p_\xi(\theta|x)=\sqrt{\xi_1}\phi(\sqrt{\xi_1}(\theta-x))$. 
Then the equilibrium cutoff in $(u,\Xi)$ is given by (\ref{regime change cutoff}), and the same discussion goes through even with heterogeneous information quality.


\subsection{Monotone payoff differentials}

The assumptions A1 and A2 require that $u(1, l, \theta)$ and $u(0, l, \theta)$ be increasing and decreasing in $(l,\theta)$, respectively.  
They guarantee the monotone comparative statics in Lemmas \ref{basic property of v} and \ref{SC}, while monotonicity of $u(1, l, \theta)-u(0, l, \theta)$ suffices in single-prior games. 
This is a limitation of our analysis because we cannot apply the main results to the class of global games with monotone payoff differentials that do not satisfy A1 and A2. 
We will be able to overcome this limitation by using recent ideas of \citet{dziewulskiquah2021}. 
 \citet{dziewulskiquah2021} introduce the notion of first-order stochastic dominance for sets of priors. 
If the set of interim beliefs with a higher private signal first-order stochastically dominates that with a lower private signal in the sense of \citet{dziewulskiquah2021}, then monotonicity of payoff differentials guarantees the monotone comparative statics in Lemmas \ref{basic property of v} and \ref{SC}. 
Such an extension of our analysis is left as a promising direction for future research.

\subsection{Smooth ambiguity preferences}

As a model with more general ambiguity-averse preferences, we can consider global games with smooth ambiguity preferences \citep{klibanoffetal2005}. 
To briefly illustrate such a model, let $\Xi=[\underline{\xi},\overline{\xi}]$ be a closed and bounded interval. 
Assume that when the opponents follow $s[\kappa]$, a player with a private signal $x$ prefers action $a$ to action $a'$ if and only if  
\begin{align}
\frac{1}{\overline{\xi}-\underline{\xi}}\int_{\underline{\xi}}^{\overline{\xi}}\phi_\alpha\left(\pi_{\xi}^a(x, \kappa)\right)d\xi \geq 
\frac{1}{\overline{\xi}-\underline{\xi}}\int_{\underline{\xi}}^{\overline{\xi}}\phi_\alpha(\pi_{\xi}^{a'}(x, \kappa))d\xi,
\notag\label{smooth preference}
\end{align}
where $\phi_\alpha(x)\equiv-\frac{1}{\alpha} e^{-\alpha x}$ with $\alpha>0$. 
The constant $\alpha$ is the coefficient of ambiguity aversion, and it is a measure of ambiguity attitude, while $\Xi$ is a measure of ambiguity perception. 
This model converges to 
the MEU model as $\alpha$ approaches infinity. 
Thus, the effects of ambiguous information studied in this paper are understood as those of ambiguity perception when the coefficient of ambiguity aversion is extremely large.  
Even if $\alpha$ is finite, we can readily show this model exhibits strategic complementarities in terms of smooth ambiguity preferences, and we can conduct a similar analysis, where the effects of ambiguous information should be weaker than those in the MEU model. 
We leave a further study of this model as a future research topic, where this paper's results serve as a benchmark.

\subsection{The assumption on the proportions of opponents' actions}
\label{SLLN sun}

One of the assumptions in players' preferences is $l_\xi(\theta)=E_\xi[\sigma|\theta]$, which follows the standard global game analysis \citep{morrisshin2002}.  
That is, when every opponent chooses a strategy $\sigma$ and the state is $\theta$, each player's evaluation of the proportion of opponents' action 1 with respect to a prior indexed by $\xi$ is equal to the expected value of $\sigma(x_j)\in\{0,1\}$ conditional on $\theta$. 
This assumption presumes the law of large numbers for a continuum of i.i.d.\ random variables, i.e., 
$\int \sigma(\theta+\varepsilon_j)dj=E_\xi[\sigma|\theta]$ almost surely, which is mathematically problematic \citep{judd1985}. 
If we consider $\{\varepsilon_i\}_{i\in [0,1]}$ as a random variable over a usual product measurable space, independence and joint measurability are incompatible, except for some trivial cases.


To elaborate on this issue, note that $l_\xi(\theta)$ is a subjective belief that the player uses to evaluate her own preferences. 
Thus, the assumption $l_\xi(\theta) = E_\xi[\sigma|\theta]$ is about players' preferences, not about an objective probabilistic law governing economic outcomes (such as the LLN in insurance markets). 
Put differently, we assume that a player subjectively believes that the LLN holds with a continuum of i.i.d.\ random variables, either naively or mathematically. 

In the latter case, players are further assumed to adopt an appropriate product probability space which guarantees the LLN.\footnote{I thank an anonymous reviewer for pointing out this issue.} 
\citet[][Proposition 5.3]{sun2006} shows the existence of a probability space that extends the usual product probability space, called the rich Fubini extension, such that independence and joint measurability are compatible and the exact LLN holds for a continuum of i.i.d. random variables with an arbitrary distribution. In our setting, a different parameter $\xi$ corresponds to a different distribution with respect to the same rich Fubini extension.

\section{Conclusion}

This paper shows that ambiguous-quality information and low-quality information have distinct effects on the equilibrium cutoffs when a safe action is not ex-ante Laplacian, on the number of equilibria when a safe action is ex-ante Laplacian, and on the likelihood of currency crises.
In our analysis, we exploit both similarities and differences between a multiple-priors game and a single-prior game. 
A multiple-priors game inherits strategic complementarities from a single-prior game, but the payoff differential alone does not determine the best responses in a multiple-priors game. 
In particular, a safe action plays a special role, e.g.,  if action 0 is a safe action, the maximum equilibrium cutoff coincides with the maximum of the maximum equilibrium cutoffs of all the fictitious single-prior games,  which leads us to our results.

We have focused on exogenously fixed ambiguity. 
However, we can extend our framework to study a social planner's decision problem who is interested in implementing desirable outcomes by engineering ambiguity, analogous to \citet{boserenou2014} and \citet{ditilloetal2017}. 
Such a problem can be referred to as ambiguous information design,\footnote{See \citet{morrisbergemann2019} for a survey on information design.} a many-player generalization of ambiguous persuasion described by \citet{beaucheneetal2019}. 
\citet{beaucheneetal2019} study a Bayesian persuasion problem \citep{kamenicagentzkow2011} with an MEU receiver, where a sender can choose an ambiguous communication device. 
One of the most successful results in information design is obtained in binary-action supermodular games \citep[e.g.,][]{inostrozapavan2020, lietal2020, morrisetal2020}, so its ambiguous version should be an important topic for future research.


\begin{appendices}
	
\bigskip
\appendix
\setcounter{section}{0}
\setcounter{theorem}{0}
\setcounter{lemma}{0}
\setcounter{claim}{0}
\setcounter{proposition}{0}
\setcounter{definition}{0}

\renewcommand{\theequation}{A.\arabic{equation} }
\setcounter{equation}{0}

\renewcommand{\thetheorem}{\Alph{theorem}}
\renewcommand{\thelemma}{\Alph{lemma}}
\renewcommand{\theclaim}{\Alph{claim}}
\renewcommand{\theproposition}{\Alph{proposition}}
\renewcommand{\thedefinition}{\Alph{definition}}

\section{Proof of Proposition \ref{lemma CvD}} 
\label{proof of lemma CvD}

We first prove Lemmas \ref{basic property of v} and \ref{SC}.

\begin{proof}[Proof of Lemma \ref{basic property of v}] 
This function is continuous by Continuity, increasing in $x$ by {Action Monotonicity}, State Monotonicity, and Stochastic Dominance, and 
decreasing in $\kappa$ by {Action Monotonicity}.
\end{proof}

\begin{proof}[Proof of Lemma \ref{SC}] 
Because $E_\xi[\sigma|\theta]\geq E_\xi[\sigma'|\theta]$, it holds that 
$u(1, E_\xi[\sigma|\theta], \theta)\geq u(1, E_\xi[\sigma'|\theta], \theta)$ and $u(0, E_\xi[\sigma|\theta], \theta)\leq u(0, E_\xi[\sigma'|\theta], \theta)$ by {Action Monotonicity}, 
which implies this lemma.
\end{proof}

We also use the following lemma, which ensures that action 0 is a dominant action when a private signal is sufficiently low, and action 1 is a dominant action when a private signal is sufficiently high.  
\begin{lemma}\label{increasing nu}
There exist $\underline{x}, \overline{x}\in \mathbb{R}$ such that 
$\min_{\xi \in \Xi}\pi_{\xi}^1(x, \kappa)-\min_{\xi \in \Xi}\pi_{\xi}^0(x, \kappa)<0$ for all 
$x\leq \underline{x}$ and $\kappa\in \mathbb{R}$ 
and 
$\min_{\xi \in \Xi}\pi_{\xi}^1(x, \kappa)-\min_{\xi \in \Xi}\pi_{\xi}^0(x, \kappa)>0$ for all $x\geq \overline{x}$ and 
 $\kappa\in \mathbb{R}$.
\end{lemma}
\begin{proof} 
We prove the existence of $\underline{x}$ (we can prove that of $\overline{x}$ similarly). 
For $\xi^0\in \arg\min_{\xi\in \Xi}\pi^0_{\xi}(x,\kappa)$, which exists by Continuity and the compactness of $\Xi$, 
\begin{align*}
\min_{\xi\in \Xi}\pi^1_{\xi}(x,\kappa)-
\min_{\xi\in \Xi}\pi^0_{\xi}(x,\kappa)
&\leq 
\pi^1_{\xi^0}(x,\kappa)-
\pi^0_{\xi^0}(x,\kappa)\leq\max_{\xi\in\Xi} E_\xi[u(1,1,\theta) -u(0,1,\theta) |x]
\end{align*}
by {Action Monotonicity}. 
Thus, it is enough to show that 
$
\lim_{x\to-\infty} \max_{\xi\in\Xi} E_\xi[u(1,1,\theta) -u(0,1,\theta) |x]<0$, 
which is true if 
\begin{equation}
\lim_{x\to-\infty} E_\xi[u(1,1,\theta) -u(0,1,\theta) |x]<0 \label{Lemma A 1}
\end{equation}
for each $\xi$ by Dini's theorem because $E_\xi[u(1,1,\theta) -u(0,1,\theta) |x]$ is increasing in $x$ by {State Monotonicity} and {Stochastic Dominance}, and $\xi\mapsto E_\xi[u(1,1,\theta) -u(0,1,\theta)|x]$ is continuous on a compact set $\Xi$. 

Let $\varepsilon=-(u(1, 1, \underline\theta)- u(0,1, \underline\theta))>0$, where $\underline{\theta}\in \mathbb{R}$ is given in {Limit Dominance}. 
Note that 
$u(1, 1, \theta)- u(0,1, \theta)\leq -\varepsilon$ for all $\theta\leq\underline\theta$ by {State Monotonicity},  and thus
\begin{align}
E_\xi[u(1,1,\theta)-u(0,1,\theta)|x]
&\leq -\varepsilon \int^{\underline{\theta}}_{-\infty} p_\xi(\theta|x)d\theta
+\int_{\underline{\theta}}^{\infty} (u(1,1,\theta) -u(0,1,\theta) )p_\xi(\theta|x)d\theta. \label{limit -1}
\end{align}
Then \eqref{Lemma A 1} holds for the following reason. 
First, by {Limit Dominance}, 
 \begin{equation}
  \lim_{x\to-\infty}\left(-\varepsilon\int^{\underline{\theta}}_{-\infty} p_\xi(\theta|x)d\theta\right)=-\varepsilon. \label{limit 1}
  \end{equation}
Next, for arbitrary $x'\in \mathbb{R}$, there exists $\hat\theta>\underline\theta$ such that $
\int_{\hat{\theta}}^{\infty} (u(1,1,\theta) -u(0,1,\theta) )p_\xi(\theta| x')d\theta<\varepsilon/2$, 
which implies that 
\[
\lim_{x\to-\infty}\int_{\hat{\theta}}^{\infty} (u(1,1,\theta) -u(0,1,\theta) )p_\xi(\theta| x)d\theta\leq\int_{\hat{\theta}}^{\infty} (u(1,1,\theta) -u(0,1,\theta) )p_\xi(\theta| x')d\theta<\varepsilon/2
\]
by {State Monotonicity} and {Stochastic Dominance}. Because 
\begin{align*}
\lim_{x\to-\infty}\left|\int_{\underline{\theta}}^{\hat\theta} (u(1,1,\theta) -u(0,1,\theta) )p_\xi(\theta|x)d\theta\right|\leq 
c\lim_{x\to-\infty}\int_{\underline{\theta}}^{\hat\theta} p_\xi(\theta|x)d\theta=0 
\end{align*}
by Limit Dominance, where $c=\max_{\theta\in [\underline{\theta},\hat\theta]}|u(1,1,\theta) -u(0,1,\theta)|$, 
we have 
\begin{equation}
\lim_{x\to-\infty}\int_{\underline{\theta}}^{\infty} (u(1,1,\theta) -u(0,1,\theta) )p_\xi(\theta|x)d\theta< \varepsilon/2. \label{limit 2}
\end{equation}
Therefore, (\ref{Lemma A 1}) holds by (\ref{limit -1}), (\ref{limit 1}), and (\ref{limit 2}). 
\end{proof} 

Using the above three lemmas, we can prove Proposition \ref{lemma CvD} by a standard argument \citep[e.g.,][]{morrisshin2002}, so we provide only a sketch of the proof (the full proof is available upon request).
Let $\Sigma_n\equiv \{\sigma\mid s[\overline{\kappa}_n](x)\leq \sigma(x)\leq s[\underline{\kappa}_n](x)\}$, where $\underline{\kappa}_0=-\infty$ and $\overline{\kappa}_0=\infty$, 
and $\underline{\kappa}_n$ and $\overline{\kappa}_n$ are defined inductively by 
\begin{align*}
\underline{\kappa}_{n+1}&=\min\{x\in \mathbb{R}\mid\min_{\xi \in \Xi}\pi_{\xi}^1(x, \underline{\kappa}_n)-\min_{\xi \in \Xi}\pi_{\xi}^0(x, \underline{\kappa}_n)
=0\},\\
\overline{\kappa}_{n+1}&=\max\{x\in \mathbb{R}\mid\min_{\xi \in \Xi}\pi_{\xi}^1(x, \overline{\kappa}_n)-\min_{\xi \in \Xi}\pi_{\xi}^0(x, \overline{\kappa}_n)=0\}.
\end{align*}
Then we can show by induction that $\Sigma_n$ is the set of strategies surviving $n$ rounds of iterated deletion of strictly interim-dominated strategies. 
Furthermore, we can show that $\underline{\kappa}=\lim_{n\to \infty}\underline{\kappa}_n $ and 
$\overline{\kappa}=\lim_{n\to \infty}\overline{\kappa}_n$, which implies Proposition \ref{lemma CvD}.


\section{Proof of Proposition \ref{main proposition}}\label{proof of main proposition}

We use the following lemma.  

\begin{lemma}\label{main lemma}
Let $f:\mathbb{R}\times \Xi \to \mathbb{R}$ be a continuous function, where $\Xi$ is a compact set. Assume that, for each $\xi\in\Xi$, $f(x, \xi)=0$ has a unique solution $x^*(\xi)$ such that $f(x, \xi)<0$ if and only if $x<x^*(\xi)$, and $x^*(\xi)$ is bounded over $\Xi$.
Then  
$\displaystyle\max_{\xi\in \Xi}x^*(\xi)$ and $\displaystyle\min_{\xi\in \Xi}x^*(\xi)$ exist, and they are unique solutions of $\displaystyle\min_{\xi\in \Xi}f(x,\xi)=0$ and $\displaystyle\max_{\xi\in \Xi}f(x,\xi)=0$, respectively. 
\end{lemma}
\begin{proof}

If $x^*(\xi)$ is continuous, 
$\displaystyle\max_{\xi\in \Xi}x^*(\xi)$ and $\displaystyle\min_{\xi\in \Xi}x^*(\xi)$ exist because $\Xi$ is compact. 
So we first show that $x^*(\xi)$ is continuous. 
Note that, for any convergent sequence $\{\xi_k\}_{k=1}^\infty$ with a limit $\bar\xi$, there exists a subsequence $\{\xi_{k_l}\}_{l=1}^\infty$ such that $\lim_{l\to \infty}x^*(\xi_{k_l})=\bar x$ for some $\bar x$ because $\{x^*(\xi_k)\}_{k=1}^\infty$ is a bounded sequence. 
Then $\lim_{l\to \infty}f(x^*(\xi_{k_l}),\xi_{k_l})=f(\bar x,\bar \xi)=0$, and thus $\bar x=x^*(\bar \xi)$. 
That is, the limit of any convergent subsequence of $\{x^*(\xi_k)\}_{k=1}^\infty$ is $\bar x$, which implies that $\lim_{k\to\infty}x^*(\xi_k)=\bar x$.

We show that $\overline{x}^*\equiv \max_{\xi\in \Xi}x^*(\xi)$ is a unique solution to $\min_{\xi\in \Xi}f(x,\xi)=0$. Note that $f(x,\xi)>0$ for all $x>\overline{x}^*\geq x(\xi)$ and $\xi\in \Xi$, which implies that 
 $\min_{\xi\in \Xi}f(x,\xi)>0$ for all $x> \overline{x}^*$ because $f(x,\xi)$ is continuous in $\xi$ and $\Xi$ is compact. 
Note also that $f(x,\xi^*)<0$ for all $x<\overline{x}^*$ if $x^*(\xi^*)=\overline{x}^*$, which implies that 
 $\min_{\xi\in \Xi}f(x,\xi)\leq f(x,\xi^*)<0$ for all $x<\overline{x}^*$. Hence, 
 $\overline{x}^*$ is the unique solution to $\min_{\xi\in \Xi}f(x,\xi)=0$ because $\min_{\xi\in \Xi}f(x,\xi)$ is continuous in $x$ (by the maximum theorem). 
We can similarly show that $\min_{\xi\in \Xi}x^*(\xi)$ is a unique solution to $\max_{\xi\in \Xi}f(x,\xi)=0$.
\end{proof}

\begin{proof}[Proof of Proposition \ref{main proposition}]
By Proposition \ref{lemma CvD}, it suffices to prove that (\ref{simple min min 2}) is a unique solution to (\ref{min min 0}). 
Let $f(\kappa,\xi_0,\xi_1)\equiv
\pi^1_{\xi_1}(\kappa, \kappa)-\pi^0_{\xi_0}(\kappa, \kappa)$.  Then (\ref{min min 0}) is written as 
$\min_{\xi_1}\max_{\xi_0}f(\kappa,\xi_0,\xi_1)=\max_{\xi_0}\min_{\xi_1}f(\kappa,\xi_0,\xi_1)=0$. 

We first show that if $\kappa<k(\xi_0, \xi_1)$, then $f(\kappa,\xi_0,\xi_1)<0$. 
Let $\kappa'< \min\{\underline{x}, \inf_{\xi_0, \xi_1}k(\xi_0, \xi_1)\}$, where $\underline{x}$ is given in Lemma~\ref{increasing nu}. 
Then it suffices to show $f(\kappa',\xi_0,\xi_1)<0$ because $k(\xi_0, \xi_1)$ is a unique solution to $f(\kappa,\xi_0,\xi_1)=0$.  
Note that 
$\max_{\xi_0}\min_{\xi_1}f(\kappa', \xi_0,\xi_1)<0$ by Lemma~\ref{increasing nu}, so there exist $\xi_0',\xi_1'\in \Xi$ such that $f(\kappa',\xi_0',\xi_1')<0$. 
Note also that $f(\kappa',\xi_0,\xi_1)\neq 0$ for all $\xi_0,\xi_1\in \Xi$ because $\kappa'< \inf_{\xi_0, \xi_1}k(\xi_0, \xi_1)$. 
Hence, we must have $f(\kappa',\xi_0,\xi_1)< 0$ for all $\xi_0,\xi_1\in \Xi$; otherwise, there would exist $\xi_0,\xi_1\in\Xi$ such that $f(\kappa',\xi_0,\xi_1)=0$ by the intermediate value theorem. 
Similarly, we can show that if $\kappa>k(\xi_0, \xi_1)$, then $f(\kappa,\xi_0,\xi_1)>0$. 

By the above argument, we can apply Lemma~\ref{main lemma} to the function $(\kappa,\xi_0)\mapsto f(\kappa,\xi_0,\xi_1)$ for each $\xi_1$; that is, $\min_{\xi_0}k(\xi_0,\xi_1)$ is the unique solution of $\max_{\xi_0}f(\kappa,\xi_0,\xi_1)=0$. 
Note that $\max_{\xi_0}f(\kappa,\xi_0,\xi_1)<0$ if and only if $\kappa<\min_{\xi_0}k(\xi_0,\xi_1)$ (see the proof of Lemma \ref{main lemma}). Thus, we can again apply Lemma~\ref{main lemma} to the function $(\kappa,\xi_1)\mapsto \max_{\xi_0}f(\kappa,\xi_0,\xi_1)$; that is, $\max_{\xi_1}\min_{\xi_0}k(\xi_0,\xi_1)$ is the unique solution of $\min_{\xi_1}\max_{\xi_0}f(\kappa,\xi_0,\xi_1)=0$. 
Similarly, 
 $\min_{\xi_0}\max_{\xi_1}k(\xi_0,\xi_1)$ is the unique solution of $\max_{\xi_0}\min_{\xi_1}f(\kappa,\xi_0,\xi_1)=\min_{\xi_1}\max_{\xi_0}f(\kappa,\xi_0,\xi_1)=0$.  
\end{proof}


\section{Proof of Proposition \ref{corollary safe} and its corollary}\label{proof of corollary safe}

\begin{proof}[Proof of Proposition \ref{corollary safe}]  
If $\kappa>\kappa^0$, then 
$\pi_\xi^1(\kappa,\kappa)-c^0(\kappa)>0$ for all $\xi\in \Xi$ because $\pi_\xi^1(\kappa,\kappa)-c^0(\kappa)\neq 0$ and $\pi_\xi^1({x},{x})-c^0({x})> 0$ for all $x>\overline{x}$, where $\overline{x}$ is given in Lemma \ref{increasing nu}. 
Thus, if $\kappa>\kappa^0$, then $
\min_{\xi}\pi_\xi^1(\kappa,\kappa)-c^0(\kappa)>0
$ by {Continuity} and the compactness of $\Xi$. 
On the other hand, $\min_{\xi}\pi_\xi^1(\kappa^0,\kappa^0)-c^0(\kappa^0)
\leq \pi_{\xi^0}^1(\kappa^0,\kappa^0)-c^0(\kappa^0)=0
$ for $
\xi^0\in \arg\max_{\xi\in\Xi} \, \max \{\kappa\mid\pi_{\xi}^1(\kappa, \kappa)=c^0(\kappa)\}$. 
Because $\min_{\xi}\pi_\xi^1(\kappa,\kappa)-c^0(\kappa)$ is continuous in $\kappa$ (by {Continuity} and the maximum theorem), $\kappa^0$ is the maximum solution to $\min_{\xi}\pi_\xi^1(\kappa,\kappa)-c^0(\kappa)=0$, i.e., the maximum equilibrium cutoff in $(u,\Xi)$. 
\end{proof} 

\begin{proof}[Proof of Corollary \ref{corollary safe 2}] 
Because $\kappa^0$ is a unique solution to $\pi_{\xi^0}^1(\kappa,\kappa)-\pi_{\xi^0}^0(\kappa,\kappa)=0$, we have
$\min_{\xi\in \Xi}\pi_\xi^1(\kappa,\kappa)-\pi_{\xi}^0(\kappa,\kappa)\leq \pi_{\xi^0}^1(\kappa,\kappa)-\pi_{\xi^0}^0(\kappa,\kappa)<0 \text{ for all $\kappa<\kappa^0$}$, which implies that $\kappa^0$ is a unique solution to $\min_{\xi}\pi_{\xi}^1(\kappa, \kappa)-c^0(\kappa)=0$.   
\end{proof}

\section{Proof of Proposition \ref{ex-ante Laplacian proposition}}\label{proof of ex-ante Laplacian proposition}

We write $q_\xi(\varepsilon_i)=\sqrt{\xi}q(\sqrt{\xi}\varepsilon_i)$. 
To prove Proposition \ref{ex-ante Laplacian proposition}, we use the following lemma.

\begin{lemma}\label{lemma for prop 4}
For any $x,\kappa\in \mathbb{R}$, if action 0 is a safe and ex-ante Laplacian action, then 
\begin{equation}
\lim_{\xi\to0}\pi_\xi^1(x,\kappa)=\lim_{\xi\to0}
\int u(1, E_\xi[s[\kappa]|\theta], \theta)p_\xi(\theta|x)d\theta
=
 \int u(1,1/2,\theta) p(\theta)d\theta<c^0. \label{uniform limit}
 \end{equation}
Thus, for any $\kappa_0,\kappa_1\in \mathbb{R}$ with $\kappa_0\leq\kappa_1$,  
there exists $\delta>0$ such that if $\xi< \delta$ then $\pi_{\xi}^1(\kappa,\kappa)\leq \pi_\xi^1(\kappa_1,\kappa_0)<c^0$ for all $\kappa\in[\kappa_0,\kappa_1]$.
\end{lemma}
\begin{proof}
It is enough to show the second equality in \eqref{uniform limit}. 
Note that 
\[
\pi_\xi^1(x,\kappa)=
\int u(1, E_\xi[s[\kappa]|\theta], \theta)p_\xi(\theta|x)d\theta
=\frac{\int u(1, E_\xi[s[\kappa]|\theta], \theta)p(\theta)q(\sqrt{\xi}(x-\theta))d\theta}{\int p(\theta)q(\sqrt{\xi}(x-\theta))d\theta}.
\]
Consider the limit of the denominator. By the monotone convergence theorem, $\int p(\theta)q(\sqrt{\xi}(x-\theta))d\theta \to \int p(\theta)q(0)d\theta=q(0)$ as $\xi\to 0$ because $q(\sqrt{\xi}(\theta-x))$ is decreasing in $\xi$ (recall the assumption that $q(\varepsilon)=q(-\varepsilon)$ and $q(\varepsilon)$ is increasing in $\varepsilon$ if $\varepsilon<0$). 
Next, consider the limit of the numerator.  
By the dominated convergence theorem, 
\[
\int u(1, E_\xi[s[\kappa]|\theta], \theta)p(\theta)q(\sqrt{\xi}(\theta-x))d\theta
\to
\int u(1, 1/2, \theta)p(\theta)q(0)d\theta
\]
as $\xi\to 0$ because 
$E_\xi[s[\kappa]|\theta]=
\int_\kappa^\infty \sqrt{\xi}q(\sqrt{\xi}(x-\theta))dx
=1- Q(\sqrt{\xi}(\kappa-\theta)) \to 1/2$, 
where $Q$ is the cumulative distribution function, 
and 
\[
|u(1, E_\xi[s[\kappa]|\theta], \theta)p(\theta)q(\sqrt{\xi}(x-\theta))|
\leq (|u(1,1,\theta)|+|u(1,0,\theta)|)p(\theta) q(0),
\]
where the right-hand side is integrable.
\end{proof}

\begin{proof}[Proof of Proposition \ref{ex-ante Laplacian proposition}]  
Fix ${\overline{\kappa}}$. 
Let $\underline{x}<{\overline{\kappa}}$ be such that $\pi_{\overline{\xi}}^1(\kappa,\kappa)<c^0= \pi_{\overline{\xi}}^0(\kappa,\kappa)$ for all $\kappa<\underline{x}$, which exists by Lemma~\ref{increasing nu}. 
Lemma \ref{lemma for prop 4} implies that   
there exists $\delta\in (0,\overline{\xi})$ such that if $\xi< \delta$ then $\pi_{\xi}^1(\kappa,\kappa)<c^0$ for all $\kappa\in[\underline{x},{\overline{\kappa}}]$.
Let $\underline{\xi}<\delta$. 
Then 
\[
\min_{\xi\in[\underline{\xi},\overline{\xi}]}\pi_{\xi}^1(\kappa,\kappa)\leq 
\begin{cases}
		\pi_{\overline{\xi}}^1(\kappa,\kappa)<c^0&\text{ if }\kappa< \underline{x},\\
		\pi_{\underline{\xi}}^1(\kappa,\kappa)<c^0&\text{ if }\kappa\in [\underline{x},{\overline{\kappa}}].
\end{cases}
\]
This implies that every equilibrium cutoff in $(u,[\underline{\xi},\overline{\xi}])$ is greater than ${\overline{\kappa}}$. 

Let $\delta$ be given above. Suppose that, for each ${\xi}<\delta$, $\pi^1_ {{\xi}}(\kappa,\kappa)=c^0$ has at most one solution greater than ${\overline{\kappa}}$. 
Let $\overline{x}>{\overline{\kappa}}$ be such that $\min_{\xi\in[\delta,\overline{\xi}]}\pi_\xi^1(\kappa,\kappa)>c^0$ for all $\kappa>\overline{x}$, which exists by Lemma~\ref{increasing nu}. 
By Lemma \ref{lemma for prop 4}, there exists $\delta'\in (0,\delta)$ such that if $\xi< \delta'$ then $\pi_{\xi}^1(\kappa,\kappa)<c^0$ for all $\kappa\in[\underline{x},\overline{x}]$.
Let $\underline{\xi}'<\delta'$. 
Then 
\begin{equation}
\min_{\xi\in[\underline{\xi}',\overline{\xi}]}\pi_{\xi}^1(\kappa,\kappa)\leq 
\begin{cases}
		\pi_{\overline{\xi}}^1(\kappa,\kappa)<c^0&\text{ if }\kappa< \underline{x},\\
		\pi_{\underline{\xi}'}^1(\kappa,\kappa)<c^0&\text{ if }\kappa\in [\underline{x},\overline{x}],
\end{cases}\label{final lemma final}
\end{equation}
so $\kappa^0=\max_{\xi\in[\underline{\xi}',\overline{\xi}]} \, \max \{\kappa\mid\pi_{\xi}^1(\kappa, \kappa)=c^0\}>\overline{x}$  because, by Proposition \ref{corollary safe}, $\kappa^0$ is the maximum solution to $\min_{\xi\in[\underline{\xi}',\overline{\xi}]}\pi_{\xi}^1(\kappa,\kappa)=c^0$.
Let $\xi^0\in \arg\max_{\xi\in[\underline{\xi}',\overline{\xi}]} \, \max \{\kappa\mid\pi_{\xi}^1(\kappa, \kappa)=c^0\}$; that is, $\pi_{\xi^0}^1(\kappa^0, \kappa^0)=c^0$. 
Then we must have $\xi^0<\delta$ because $\kappa^0>\overline{x}$ implies $\min_{\xi\in[\delta,\overline{\xi}]}\pi_\xi^1(\kappa^0,\kappa^0)>c^0$. 
Thus, $\pi_{\xi^0}^1({{\kappa}}, {{\kappa}})< c^0$ for all $\kappa\in[\underline{x},{\overline{\kappa}}]$ by the definition of $\delta$. 
Note that $\pi_{\xi^0}^1(\kappa, \kappa)\neq c^0$ for all $\kappa\in ({\overline{\kappa}}, \kappa^0)$ by the assumption. 
Accordingly, $\min_{\xi\in[\underline{\xi}',\overline{\xi}]}\pi_{\xi}^1(\kappa,\kappa)\leq \pi_{\xi^0}^1(\kappa, \kappa)< c^0$ for all $\kappa\in ({\overline{\kappa}},\kappa^0)$ by Continuity. 
This implies that
$\min_{\xi\in[\underline{\xi}',\overline{\xi}]}\pi_{\xi}^1(\kappa,\kappa)< c^0$ for all $\kappa<\kappa^0$ by \eqref{final lemma final}.  
Therefore, by Proposition \ref{corollary safe}, $s[\kappa^0]$ is the unique strategy surviving iterated deletion of strictly interim-dominated strategies in $(u,[\underline{\xi}',\overline{\xi}])$.
\end{proof}

\end{appendices}


\begin{thebibliography}{99}

\bibitem[Angeletos et al.(2007)]{angeletosetal2007}
Angeletos, G.\ M., Hellwig, C., Pavan, A., 2007. Dynamic global games of regime change: learning, multiplicity, and the timing of attacks. Econometrica 75, 711--756.

\bibitem[Angeletos and Lian(2016)]{angeletoslian2016}
Angeletos, G.\ M., Lian, C., 2016. 
Incomplete information in macroeconomics: accommodating frictions in coordination. 
In: Taylor, J.\ B., Uhlig, H.\ (Eds.), Handbook of Macroeconomics, vol.\ 2B. Elsevier, pp.\ 1345-1425.










\bibitem[Ahn(2007)]{ahn2007}
Ahn, D., 2007. Hierarchies of ambiguous beliefs. J. Econ. Theory 136, 286--301.

\bibitem[Azrieli and Teper(2011)]{azrieliteper2011}
Azrieli, Y., Teper, R., 2011. 
Uncertainty aversion and equilibrium existence in games with incomplete information. 
Games Econ.\ Behav.\ 73, 310--317.



\bibitem[Aumann(1976)]{aumann1976}Aumann, R. J., 1976. Agreeing to disagree. Ann. Statist. 4, 1236--1239.







\bibitem[Beauch\^ene et al.(2019)]{beaucheneetal2019}
Beauch\^ene, D., Li, J., Li, M., 2019. 
Ambiguous persuasion. J.\ Econ.\ Theory 179, .312--365.


\bibitem[Bergemann and Morris(2019)]{morrisbergemann2019} Bergemann, D., Morris, S., 2019.  
Information design: A unified perspective. J.\ Econ.\ Lit.\ 57, 44--95.  

\bibitem[Bodoh-Creed(2012)]{bodohcreed2012}
Bodoh-Creed, A. L., 2012. 
Ambiguous beliefs and mechanism design. Games Econ.\ Behav.\ 75 518--537.



\bibitem[Bose and Daripa(2009)]{bosedaripa2009}
Bose, S., Daripa, A., 2009. 
A dynamic mechanism and surplus extraction under ambiguity. 
J.\ Econ.\ Theory 144, 2084--2114.

\bibitem[Bose et al.(2006)]{boseetal2006}
Bose, S., Ozdenoren, E., Pape, A., 2006. 
Optimal auctions with ambiguity. 
Theoretical Econ. {1}, 411--438. 

\bibitem[Bose and Renou(2014)]{boserenou2014}
Bose, S., Renou, L., 2014. 
Mechanism design with ambiguous communication devices. 
Econometrica 82, 1853--1872.





\bibitem[Caballero and Krishnamurthy(2008)]{caballerokrishnamurthy2008}
Caballero, R. J., Krishnamurthy, A., 2008. 
Collective risk management in a flight to quality episode. 
J.\ Finance 63, 2195--2230.

\bibitem[Calvo(1988)]{calvo1988}
Calvo, G.\ A., 1988. Servicing the public debt: the role of expectations. Am. Econ. Rev. 78, 647--661.



\bibitem[Carlsson and van Damme(1993)]{carlssonvandamme1993a} 
Carlsson, H., van Damme, E., 1993.
Global games and equilibrium selection. 
Econometrica 61, 989--1018.










\bibitem[De Castro and Yannelis(2018)]{decastroyannelis2018}
De Castro, L., Yannelis, N.\ C., 2018. Uncertainty, efficiency and incentive compatibility: ambiguity solves the conflict between efficiency and incentive compatibility. J. Econ. Theory 177, 678--707.

\bibitem[Diamond and Dybvig(1983)]{diamonddybvig1983} 
Diamond, D., Dybvig, P., 1983. 
Bank runs, deposit insurance, and liquidity. 
J. Polit. Economy 91, 401--419. 


\bibitem[Di Tillio et al.(2017)]{ditilloetal2017}
Di Tillio, A., Kos, N., Messner, M., 2017. 
The design of ambiguous mechanisms. Rev.\ Econ.\ Stud.\ 84, 237--276.

\bibitem[Dicks and Fulghieri(2019)]{dicksfulghieri2019}
Dicks, D., Fulghieri, P., 2019. Uncertainty aversion and systemic risk. J.\ Political Econ.\ 127, 1118--1155.


\bibitem[Dziewulski and Quah(2024)]{dziewulskiquah2021} 
Dziewulski, P., Quah, J.\ K.-H., 2024. Comparative statics with linear objectives: Normality, complementarity, and ranking multi-prior beliefs. Econometrica 92, 167--200.








\bibitem[Ellis(2016)]{ellis2016}
Ellis, A., 2016. Condorcet meets Ellsberg. Theoretical Econ.\ 11, 865--895.





\bibitem[Epstein(1997)]{epstein1997}
Epstein, L. G., 1997.  Preference, rationalizability and equilibrium. J.\ Econ.\ Theory 73, 1--29.










\bibitem[Epstein and Wang(1996)]{epsteinwang1996}
Epstein, L. G., Wang, T., 1996. 
``Beliefs about beliefs'' without probabilities. 
Econometrica 64, 1343--1373. 

\bibitem[Fabrizi et al.(2022)]{fabrizi2019}
Fabrizi, S., Lippert, S., Pan, A., Ryan, M., 2022. 
A theory of unanimous jury voting with an ambiguous likelihood. Theory Dec.\ 93, 399--425. 


\bibitem[Fagin and Halpern(1990)]{faginhalpern1990}
Fagin, R., Halpern, J., 1990. 
A new approach to updating beliefs.  
In: Bonissone, P. P., Henrion, M.,  Kanal, L. N., Lemmer, J. F. (Eds.),  
Uncertainty in Artificial Intelligence~6. 
Elsevier, pp.\ 317--325. 






\bibitem[Gilboa and Marinacci(2013)]{gilboamarinacci2013}
Gilboa, I., Marinacci, M., 2013. 
Ambiguity and the Bayesian paradigm.
In: Acemoglu, D., Arellano, M.,  Dekel, E. (Eds.),  
Advances in Economics and Econometrics: Tenth World Congress vol.\ 1. 
Cambridge Univ.\ Press, pp.\ 179--242. 









\bibitem[Gilboa and Schmeidler(1989)]{gilboaschmeidler1989}
Gilboa, I., Schmeidler, D., 1989. Maxmin expected utility with a non-unique prior. J.
Math. Econ. 18, 141--153.



\bibitem[Goldstein and Pauzner(2005)]{goldsteinpauzner2005}
Goldstein, I., Pauzner, A., 2005. 
Demand-deposit contracts and the probability of bank runs. 
J. Finance 60, 1293--1327. 

\bibitem[Grant et al.(2016)]{gratetal2016}
Grant, S., Meneghel, I.,  Tourky, R., 2016. Savage games. Theoretical Econ.\ 11, 641--682. 

 


\bibitem[Guo and Yannelis(2021)]{guoyannelis2021}
Guo, H., Yannelis, N.\ C., 2021. 
Full implementation under ambiguity. Am.\ Econ.\ J.\ Microecon.\  13, 148--178.


 
\bibitem[Hanany et al.(2020)]{hananyetal2020}
Hanany, E., Klibanoff, P., Mukerji, S., 2020. Incomplete information games with ambiguity averse players. Amer.\ Econ.\  J.\ Microecon.\  12, 135--187.
 
\bibitem[Harsanyi(1967--1968)]{harsanyi1967}
Harsanyi, J. C., 1967--1968. Games with incomplete information played by Bayesian players. 
Manage. Sci. 14, 159--182, 320--334, 486--502. 











\bibitem[Iachan and Nenov(2015)]{iachannenov2015}
Iachan, F., Nenov, P., 2015. Information quality and crises in regime-change games.\ J.\ Econ.\ Theory 158, 739--768.

\bibitem[Inostroza and Pavan(2022)]{inostrozapavan2020}
Inostroza, N., Pavan, A., 2022. 
Adversarial coordination and public information design. Working paper.

\bibitem[Jaffray(1992)]{jaffray1992}
Jaffray, J.-Y., 1992. Bayesian updating and belief functions.  
IEEE Trans.  Syst. Man Cybern. 22, 1144--1152. 


\bibitem[Judd(1985)]{judd1985}
Judd, K. L., 1985. The law of large numbers with a continuum of IID random variables. J.\ Econ.\ Theory 35, 19--25.




\bibitem[Kajii and Ui(2005)]{kajiiui2005}
Kajii, A., Ui, T., 2005.  Incomplete information games with multiple priors. {Japanese Econ. 
Rev.} 56, 332--351.

\bibitem[Kajii and Ui(2009)]{kajiiui2009}
Kajii, A., Ui, T., 2009.  Interim efficient allocations under uncertainty. J. Econ. Theory 144, 337--353. 




\bibitem[Kamenica and Gentzkow(2011)]{kamenicagentzkow2011}
Kamenica, E., Gentzkow, M., 2011. Bayesian persuasion. 
 Amer.\ Econ.\ Rev. 101, 2590--2615.
 

\bibitem[Kawagoe and Ui(2013)]{kawagoeui2013}
Kawagoe, T., Ui, T., 2013.  Global games and ambiguous information: an experimental study. Working paper. 



\bibitem[Klibanoff et al.(2005)]{klibanoffetal2005}
Klibanoff, P., Marinacci, M., Mukerji,  S., 2005. 
A smooth model of decision making under ambiguity. 
Econometrica 73, 1849--1892.



\bibitem[Laskar(2014)]{laskar2014}
Laskar, D., 2014. 
Ambiguity and perceived coordination in a global game. 
Econ.\ Letters 122, 317--320. 

\bibitem[Li et al.(2023)]{lietal2020}
Li, F., Song, Y., Zhao, M., 2023. 
Global manipulation by local obfuscation. 
J. Econ.\ Theory 207, 105575.


\bibitem[Lipsky(1998)]{lipsky1998}
Lipsky, J., 1998. Asia's crisis: a market perspective. Finance Dev. 35, 10--13.



\bibitem[Lo(1998)]{lo1998}
Lo, K.\ C., 1998. Sealed bid auctions with uncertainty averse bidders. 
Econ. Theory 12, 1--20. 





\bibitem[Machina and Siniscalchi(2014)]{machinasiniscalchi2014}
Machina, M., Siniscalchi, M., 2014. 
Ambiguity and ambiguity aversion. 
In: Machina, M., Viscusi, K.\ (Eds.), 
Handbook of the Economics of Risk and Uncertainty, vol.\ 1. Elsevier, pp.\ 729--807.




\bibitem[Martins-da-Rocha(2010)]{martinsdarocha2010}
Martins-da-Rocha, V.\ F., 2010. 
Interim efficiency with MEU-preferences. 
J.\ Econ.\ Theory 145, 1987--2017. 
  
  
  
\bibitem[Mertens and Zamir(1985)]{mertenszamir1985}
Mertens, J.-F., Zamir, S., 1985. 
Formulation of Bayesian analysis for games with incomplete information. 
Int. J. Game Theory 14, 1--29. 


\bibitem[Milgrom(1981)]{milgrom1981}  
Milgrom, P., 1981.
Good news and bad news: representation theorems and applications. 
Bell J. Econ.\ 12, 380--391. 

\bibitem[Milgrom and Roberts(1990)]{milgromroberts1990}
Milgrom, P., Roberts, J., 1990. 
Rationalizability, learning, and equilibrium in games with strategic complementarities. 
Econometrica 58, 1255--1277.











\bibitem[Morris and Shin(1998)]{morrisshin1998} Morris, S., Shin, H. S., 
1998.  
Unique equilibrium in a model of self-fulfilling currency attacks.
Am. Econ. Rev. 88, 587--597. 


\bibitem[Morris and Shin(2001)]{morrisshin2001} Morris, S., Shin, H. S., 2001.  Rethinking multiple equilibria in macroeconomic modeling.  NBER Macroeconomics Annual 2000, 139--161. 

\bibitem[Morris and Shin(2003)]{morrisshin2002} Morris, S., Shin, H. S., 2003.  Global games: theory and applications.  
In: Dewatripont, M., Hansen, L., Turnovsky, S. (Eds.), 
Advances in Economics and Econometrics: Theory and Applications, 
Proceedings of the Eighth World Congress of the Econometric Society.  
Cambridge Univ.\ Press, pp.\ 56--114. 

\bibitem[Morris and Shin(2004)]{morrisshin2004} Morris, S., Shin, H. S., 2004. Coordination risk and the price of debt.  Eur. Econ. Rev. 48, 133--153. 

\bibitem[Morris et al.(2024)]{morrisetal2020} Morris, S., Oyama, D., Takahashi, S., 2024. Implementation via information design in binary-action supermodular games. Econometrica 92, 775--813.



\bibitem[Obstfeld(1996)]{obstfeld1996} 
Obstfeld, M., 1986. Models of currency crises with self-fulfilling features. Eur. Econ. Rev. 40, 1037--1047.
 
 






\bibitem[Pan(2019)]{pan2019}
Pan, A., 2019. A note on pivotality. Games 10, 24.



\bibitem[Pires(2002)]{pires2002}
Pires, C. P., 2002. A rule for updating ambiguous beliefs.  
Theory Dec. 53, 1573--7187. 





\bibitem[Routledge and Zin(2009)]{routledgezin2009}  
Routledge, B, Zin, S., 2009. 
Model uncertainty and liquidity. Rev.\ Econ.\ Dynamics 12, 543


\bibitem[Ryan(2021)]{ryan2021}
Ryan, M., 2021. 
Feddersen and Pesendorfer meet Ellsberg. 
Theory Dec. 90, 543--577.
 
 
\bibitem[Salo and Weber(1995)]{saloweber1995}
Salo, A., Weber, M., 1995. 
Ambiguity aversion in first-price sealed-bid auctions.  
J. Risk Uncertainty 11, 123--137. 







\bibitem[Siamwalla(2005)]{siamwalla2005}
Siamwalla, A., 2005. Anatomy of the crisis. In: Warr,  P. (Ed.), Thailand beyond the Crisis. Routledge Curzon, pp. 66--104.



\bibitem[Song(2018)]{song2018}
Song, Y., 2018. Efficient implementation with interdependent valuations and maxmin agents. J. Econ. Theory 176, 693--726.









\bibitem[Sun(2006)]{sun2006} 
Sun, Y, 2006. The exact law of large numbers via Fubini extension and characterization of insurable risks. J.\ Econ.\ Theory 126, 31--69.



\bibitem[Uhlig(2010)]{uhlig2010} 
Uhlig, H., 2010. A model of a systemic bank run. J.\ Monetary Econ.\ 57, 78--96.

\bibitem[Ui(2009)]{ui2009} 
Ui, T., 2009. 
Ambiguity and risk in global games. Working paper. 


\bibitem[Van Zandt and Vives(2007)]{vanzandtvives2007}
Van Zandt, T., Vives, X., 2007. Monotone equilibria in Bayesian games of strategic complementarities.
J.\ Econ.\ Theory 134, 339--360.


\bibitem[Vives(1990)]{vives1990} 
Vives, X., 1990.  
Nash equilibrium with strategic complementarities. 
J.\ Math. Econ. 19, 305--321.




\bibitem[Wolitzky(2016)]{wolitzky2016}
Wolitzky, A., 2016. Mechanism design with maxmin agents: theory and an application to bilateral trade. Theor.\ Econ.\ 11, 971--1004.





\end{thebibliography}
\end{document}